\documentclass[aps, prd, twocolumn, superscriptaddress, nofootinbib]{revtex4-1}

\usepackage{bm,amssymb,slashed,graphicx,multirow,soul,mathtools,xspace,array}

\usepackage{hyperref}

\newcommand{\dds}{D^{(*)}}
\newcommand{\dss}{D^{**}}
\newcommand{\dsW}{D^{1/2^+}}
\newcommand{\dsN}{D^{3/2^+}}
\newcommand{\dSs}{D^*_0}
\newcommand{\dVs}{D^*_1}
\newcommand{\dV}{D_1}
\newcommand{\dTs}{D^*_2}
\newcommand{\bbar}{\bar{b}}
\newcommand{\cbar}{\bar{c}}
\newcommand{\Bbar}{\bar{B}}
\newcommand{\Dbar}{\bar{D}}

\newcommand{\mn}{{\mu\nu}}
\newcommand{\g}{\gamma}
\newcommand{\ampB}[2]{\big\langle #1 \big|\, #2\, \big| \Bbar \big\rangle }
\newcommand{\ampBs}[2]{\langle #1 |\, #2\, | \Bbar \rangle}

\def\spnt{1}
\if\spnt1

\else

\fi	
\newcommand{\ov}{\overline}
\newcommand{\aS}{\alpha_s}
\newcommand{\haS}{{\hat{\alpha}_s}}
\newcommand{\nn}{\nonumber}
\newcommand{\GeV}{\text{GeV}}
\newcommand{\MeV}{\text{MeV}}
	
\def\lqcd{\Lambda_\text{QCD}}

\newcommand{\beq}{\begin{equation}}
\newcommand{\eeq}{\end{equation}}
\newcommand{\beqa}{\begin{eqnarray}}
\newcommand{\eeqa}{\end{eqnarray}}

\newcommand{\alSL}{\alpha_L^S}
\newcommand{\alSR}{\alpha_R^S}
\newcommand{\alVL}{\alpha_L^V}
\newcommand{\alVR}{\alpha_R^V}
\newcommand{\alTL}{\alpha_L^T}
\newcommand{\alTR}{\alpha_R^T}
\newcommand{\beSL}{\beta_L^S}
\newcommand{\beSR}{\beta_R^S}
\newcommand{\beVL}{\beta_L^V}
\newcommand{\beVR}{\beta_R^V}
\newcommand{\beTL}{\beta_L^T}
\newcommand{\beTR}{\beta_R^T}
\newcommand{\rV}{}
\newcommand{\rS}{}
\newcommand{\rT}{}
\newcommand{\rC}{r}

\newcommand{\tA}{\tilde\alpha}
\newcommand{\tB}{\tilde\beta}
\newcommand{\alSLt}{\tA_L^S}
\newcommand{\alSRt}{\tA_R^S}
\newcommand{\alVLt}{\tA_L^V}
\newcommand{\alVRt}{\tA_R^V}
\newcommand{\alTLt}{\tA_L^T}
\newcommand{\alTRt}{\tA_R^T}
\newcommand{\beSLt}{\tB_L^S}
\newcommand{\beSRt}{\tB_R^S}
\newcommand{\beVLt}{\tB_L^V}
\newcommand{\beVRt}{\tB_R^V}
\newcommand{\beTLt}{\tB_L^T}
\newcommand{\beTRt}{\tB_R^T}

\newcommand{\ds}{\displaystyle}

\def\d{{\rm d}}
\newcommand{\rl}{\rho_\ell}
\newcommand{\rt}{r_\ell}
\newcommand{\mSqq}{\hat q^2}
\newcommand{\thtau}{\theta_{\ell}}
\newcommand{\phtau}{\phi_{\ell}}

\newcommand{\phD}{\phi_{D}}
\newcommand{\phDtau}{(\phD - \phtau)}

\makeatletter
\g@addto@macro\bfseries{\boldmath}
\makeatother

\allowdisplaybreaks

\graphicspath{{./DssFigures/}}

\begin{document}

\title{Model independent analysis of semileptonic $B$ decays to $\dss$ for
arbitrary new physics}

\author{Florian U.\ Bernlochner}
\affiliation{Physikalisches Institut der Rheinischen Friedrich-Wilhelms-Universit\"at Bonn, 53115 Bonn, Germany}
\affiliation{Karlsruher Institute of Technology, 76131 Karlsruhe, Germany}

\author{Zoltan Ligeti}
\affiliation{Ernest Orlando Lawrence Berkeley National Laboratory, 
University of California, Berkeley, CA 94720, USA}

\author{Dean J.\ Robinson}
\affiliation{Physics Department, University of Cincinnati,
Cincinnati OH 45221, USA}

\begin{abstract}

We explore semileptonic $B$ decays to the four lightest excited charm mesons,
$D^{**} = \{D_0^*,\, D_1^*,$ $D_1,\, D_2^*\}$, for nonzero charged lepton mass and
for all $b\to c \ell\bar\nu$ four-Fermi interactions, including calculation of
the ${\cal O}(\Lambda_\text{QCD}/m_{c,b})$ and ${\cal O}(\alpha_s)$ corrections to the heavy
quark limit for all form factors.  In the heavy quark limit some form factors
are suppressed at zero recoil, therefore, the ${\cal O}(\Lambda_\text{QCD}/m_{c,b})$
corrections can be very important.  The $D^{**}$ rates exhibit sensitivities to
new physics in $b\to c\tau\bar\nu$ mediated decays complementary to the $D$ and
$D^*$ modes.  Since they are also important backgrounds to $B\to
D^{(*)}\tau\bar\nu$, the correct interpretation of future semitauonic $B\to
D^{(*)}$ rate measurements requires consistent treatment of both the $D^{**}$
backgrounds and the signals.  Our results allow more precise and more reliable
calculations of these $B\to D^{**}\ell\bar\nu$ decays, and are systematically
improvable by better data on the $e$ and $\mu$ modes. As an example, we show
that the $D^{**}$ rates are more sensitive to a new $\bar c\,
\sigma_{\mu\nu} b$ tensor interaction than the $D^{(*)}$ rates.

\end{abstract}

\maketitle

\section{Introduction}

The measurements of the ratio of semitauonic $B$ decays
compared to the light-lepton final states, 
\beq\label{RXdef}
	R(X) = \frac{\Gamma(B\to X\tau\bar\nu)}{\Gamma(B\to X l\bar\nu)}\,, 
  \qquad l = \mu, \,e\,,
\eeq
show a $4 \sigma$ tension with the standard model (SM) expectations~\cite{HFAG},
when the $X=D$ and $D^*$ results are combined.  Improving our understanding of
the $B \to D^{(*)}$ form factors, required for precision calculations of
$R(\dds)$, has received renewed attention recently~\cite{Bernlochner:2017jka,
Bigi:2017njr, Grinstein:2017nlq, Bernlochner:2017xyx, Lattice:2015rga,
BDsLatticeAllw, Jaiswal:2017rve}.  To maximize future sensitivity to new physics
(NP) contributions, measuring and understanding contributions for additional
semileptonic decay modes mediated by the same parton-level transition is
important and necessary, not only as they can give complementary information on
the new physics, but also as they constitute backgrounds to the $R(\dds)$
measurements.

In this paper we study $B \to \dss\ell\bar\nu$ decays, where
\beq\label{Dss}
	\dss \in \big\{D_0^*,\, D_1^*,\, D_1,\, D_2^*\big\}\,,
\eeq
denotes the four lightest excited charmed mesons, above the $\{D, D^*\}$
ground-state doublet of heavy quark symmetry (HQS)~\cite{Isgur:1989vq,
Isgur:1989ed, Isgur:1991wq}.  (The $\dss$ notation is common in the experimental
literature; these are the $1P$ orbitally excited states in the quark model.)  In
Ref.~\cite{Bernlochner:2016bci}, SM predictions for $R(\dss)$ were derived,
extending results for massless leptons~\cite{Leibovich:1997tu,
Leibovich:1997em}, but a comprehensive study of NP effects has not been carried
out yet.  We include contributions from all possible four-fermion
operators (assuming no right-handed neutrinos), and derive the
$\mathcal{O}(\lqcd/m_{c,b})$ and $\mathcal{O}(\aS)$ terms in the expansions of
the form factors, going beyond the leading order in the heavy quark expansion.
The ${\cal O}(\lqcd/m_{c,b})$ corrections to the SM matrix elements were
calculated a long time ago~\cite{Leibovich:1997tu, Leibovich:1997em}, and can be
substantial, due to the suppressions of certain leading order matrix elements
near zero recoil, imposed by heavy quark symmetry.
We show that the available $B\to D_2^*l\bar\nu$ and $B\to D_1l\bar\nu$ data are
in severe tension with the heavy quark limit, that is alleviated by including
$\mathcal{O}(\lqcd/m_{c,b})$ corrections.
Similarly, $\mathcal{O}(\lqcd/m_{c,b})$ terms can generate numerically dominant
contributions to NP matrix elements as well, and must be included.

Understanding the $B \to \dss\ell\bar\nu$ decays as precisely as possible, both
theoretically and experimentally, is important for several reasons. First, as in
$B \to \dds\ell\bar\nu$ decays, certain form factor combinations are suppressed
by the light lepton mass, and thus cannot be constrained by $B \to \dss l \nu$
measurements, but enter unsuppressed in the semitauonic rates. The use of heavy
quark effective  theory (HQET)~\cite{Georgi:1990um, Eichten:1989zv} allows more
precise future measurements of $B\to \dss l\bar\nu$ to systematically improve
the predictions for $B\to \dss \tau\bar\nu$~\cite{Bernlochner:2016bci}, which
will provide complementary sensitivity to new physics compared to $B \to
\dds\tau\bar\nu$. Second, $B \to \dss\ell\bar\nu$ decays also constitute a
significant background to the measurements of $R(\dds)$, contributing
significantly to its uncertainty at present.  As certain $B \to \dss\tau\bar\nu$
modes may exhibit high sensitivity to NP, good theoretical control of these
backgrounds is required in order to understand which NP operators may best fit
the data.  Third, better theoretical control of these modes will help to improve
the determinations of the CKM elements $|V_{cb}|$ and $|V_{ub}|$, both from
exclusive and inclusive $B$ decays.  The study of these decay
modes~\cite{Isgur:1990jf} and their contributions to the Bjorken sum
rule~\cite{Bjorken:1990hs} will help understanding the composition of the
inclusive $B\to X_c \ell\bar\nu$ decay in terms of exclusive modes.

In Sec.~\ref{sec:hqet} we establish notations and calculate all $B\to \dss$ form
factors, including the complete set of order $\mathcal{O}(\lqcd/m_{c,b})$ and
$\mathcal{O}(\alpha_s)$ effects. Section~\ref{sec:nprates} contains expressions
for the differential decay rates for arbitrary currents and charged lepton
mass.  In Sec.~\ref{sec:app} we study observables that are particularly
sensitive to the ${\cal O}(\lqcd/m_{c,b})$ corrections, and plot effects of a NP
tensor interaction which could not be evaluated previously with comparable
accuracy.  Section~\ref{sec:concl} concludes.

\section{HQET expansion of the form factors}
\label{sec:hqet}

We are interested in the $\Bbar\to \dss$ matrix elements of operators with all
possible Dirac structures, for which we choose the basis
\begin{align}\label{eqn:Odef}
	O_V &= \bar c\,\g_\mu\, b\,, & O_A &= \bar c\, \g_\mu\g_5\, b\,, && \nn\\*	
	O_S &= \bar c\, b\,, &  O_P & = \bar c\, \g_5\, b\,, & O_T & = \bar c\, \sigma_{\mu\nu}\, b\,,
\end{align}
with $\sigma_{\mu\nu} = (i/2)\, [\g_\mu,\g_\nu]$.  Throughout this paper we
assume isospin symmetry, and $\Bbar$ denotes $\Bbar^0$ or $B^-$.  As in
Refs.~\cite{Leibovich:1997tu, Leibovich:1997em, Bernlochner:2016bci}, we use
the conventions $\text{Tr}[\g^\mu\g^\nu\g^\sigma\g^\rho\g^5] = -4i
\epsilon^{\mu\nu\rho\sigma}$, so that $\sigma^{\mu\nu} \g^5 \equiv
+(i/2)\epsilon^{\mu \nu \rho \sigma} \sigma_{\rho \sigma}$. 
(This is the opposite of the common
convention in the $\Bbar \to D^{(*)}\ell\bar\nu$ literature, which typically
chooses $\text{Tr}[\g^\mu\g^\nu\g^\sigma\g^\rho\g^5] = +4i
\epsilon^{\mu\nu\rho\sigma}$, so that $\sigma^{\mu\nu} \g^5 \equiv
-(i/2)\epsilon^{\mu \nu \rho \sigma} \sigma_{\rho \sigma}$.)

\subsection{Spectroscopy}

The spectroscopy of the $\dss$ states is important, because in addition to the
impact on the kinematics, it also affects the HQET expansion of the form
factors~\cite{Leibovich:1997tu, Leibovich:1997em}.  The isospin averaged masses
and widths for the six lightest charm meson states are shown in
Table~\ref{tab:charm}.  (The level of agreement between the measurements of the
masses and widths of the $\dss$ mesons, especially those of $\dSs$ in the top
row of Table~\ref{tab:charm}, is  presently
unsatisfactory~\cite{Bernlochner:2016bci}.)


\begin{table}[tb]
\renewcommand*{\arraystretch}{1.2}
\tabcolsep 6pt
\begin{tabular}{ccccc}
\hline\hline
Particle  &    $s_l^{\pi_l}$ &  $J^P$  &  $m$ (MeV)  &  $\Gamma$ (MeV)\\
\hline
$\dSs$ &  $\frac12^+$  &  $0^+$  &  $2349$  &  $236$ \\
$\dVs$ &  $\frac12^+$  &  $1^+$  &  $2427$  &  $384$ \\
\hline
$\dV$ &  $\frac32^+$  &  $1^+$  &  $2421$  &  $31$ \\
$\dTs$ &  $\frac32^+$  &  $2^+$  &  $2461$  &  $47$ \\
\hline\hline
$D^*$ &  $\frac12^-$  &  $1^-$  &  $2009$  &  0. \\
$D$ &  $\frac12^-$  &  $0^-$  &  $1866$  &  0. \\
\hline\hline
\end{tabular}
\caption{Isospin averaged masses and widths of the six lightest charm mesons,
rounded to 1\,MeV~\cite{[][; and updates at \url{http://pdglive.lbl.gov/}]PDG}.}
\label{tab:charm}
\end{table}

In the heavy quark limit, the spin-parity of the
light degrees of freedom, $s_l^{\pi_l}$, is a conserved quantum number,
yielding doublets of heavy quark symmetry, as the spin $s_l$ is combined with the
heavy quark spin~\cite{Isgur:1991wq}.  In the quark model, the four $\dss$
states correspond to combining the heavy quark and light quark spins with
$L=1$ orbital angular momentum.  The masses of each heavy quark spin symmetry doublet of hadrons, $H_\pm$,
with total spin $J_\pm = s_l \pm \frac12$ can be expressed in HQET as
\begin{equation}\label{mass}
m_{H_\pm} = m_Q + \bar\Lambda^H - {\lambda_1^H \over 2 m_Q} 
  \pm {n_\mp\, \lambda_2^H \over 2m_Q} + \ldots \,,
\end{equation}
where $n_\pm = 2J_\pm+1$ is the number of spin states of each hadron, and the
ellipsis denote terms suppressed by more powers of  $\Lambda_{\rm QCD}/m_Q$. 
The parameter $\bar\Lambda^H$ is the energy of the light degrees of freedom
in the $m_Q\to\infty$ limit, and plays an important role, as it is related to
the semileptonic form factors~\cite{Leibovich:1997tu, Leibovich:1997em}.  
The $\lambda_1^H$ and $\lambda_2^H$ parameters are related to the heavy quark
kinetic energy and chromomagnetic energy in the hadron~$H$. We use the
notation $\bar\Lambda$, $\bar\Lambda'$, and $\bar\Lambda^*$ for the
$\frac12^-$, $\frac32^+$, and $\frac12^+$ doublets, respectively, and for the
states in each doublet
\beq\label{sad}
\dsW \in \big\{D_0^*,\, D_1^*\big\}\,, \qquad
  \dsN \in \big\{D_1,\, D_2^*\big\}\,.
\eeq

The current data suggest that the $m_{D_1^*} - m_{D_0^*}$ mass splitting is
substantially larger than $m_{D_2^*} - m_{D_1}$.  This possibility was not
considered in Refs.~\cite{Leibovich:1997tu, Leibovich:1997em}, since at that
time both of these mass splittings were about 40\,MeV.  The smallness of
$m_{D_2^*} - m_{D_1}$ and $m_{D_1^*} - m_{D_0^*}$ compared to $m_{D^*} - m_D
\simeq 140\,\MeV$ was taken as an indication that the chromomagnetic operator
matrix elements are suppressed for the four $\dss$ states, in agreement with
quark model predictions.  We relax this constraint, as in
Ref.~\cite{Bernlochner:2016bci}.

\begin{table}[t]
\newcolumntype{C}{ >{\centering\arraybackslash } m{0.8cm} <{}}
\renewcommand*{\arraystretch}{1.3}
\begin{tabular}{c|CCC|CCC}
\hline\hline
Parameter  &  $\bar\Lambda$  &  $\bar\Lambda'$  &  $\bar \Lambda^*$ 
  &  $\bar\Lambda_s$  &  $\bar\Lambda'_{s}$  &  $\bar \Lambda^*_{s}$ \\
  \hline
Value [GeV]  &  0.40 & 0.80 & 0.76  &  0.49 & 0.90 & 0.77 \\
\hline\hline
\end{tabular}
\caption{The HQET parameter estimates used~\cite{Bernlochner:2016bci}.}
\label{tab:input_summary}
\end{table}

While the measured masses of the broad $D_0^*$ and $D_1^*$ states changed
substantially over the last twenty years, their $2J+1$ weighted average is
essentially unchanged compared to Ref.~\cite{Leibovich:1997em}.  We use
$\bar\Lambda' - \bar\Lambda = 0.40\,\GeV$ and $\bar\Lambda' - \bar\Lambda^*
\simeq 0.04\,\GeV$, and summarize the parameters used in
Table~\ref{tab:input_summary}.  The uncertainty of $\bar\Lambda$ is
substantially greater than that of $\bar\Lambda' - \bar\Lambda$ and
$\bar\Lambda' - \bar\Lambda^*$; as we see below, the form factors are less
sensitive to $\bar\Lambda$ than to these splittings.

\subsection{Matrix elements to order \texorpdfstring{$\lqcd/m_{c,b}$}{lqcdm} and \texorpdfstring{$\aS$}{as}}
\label{sec:form}

It is simplest to calculate the $\Bbar \to \dss$ matrix elements in HQET using the
trace formalism~\cite{Falk:1990yz, Bjorken:1990rr, Falk:1991nq}.  It allows a
straightforward evaluation of the matrix elements of the five operators in
Eq.~\eqref{eqn:Odef}, as well as those of additional operators generated by
perturbative corrections (for a review, see
Ref.~\cite{Manohar:2000dt}). The ${\cal O}(\aS)$ corrections are given
explicitly in Appendix~A of Ref.~\cite{Bernlochner:2017jka}, extracted from
Refs.~\cite{Falk:1990yz, Falk:1990cz, Neubert:1992qq}.  The ${\cal
O}(\alpha_s\, \lqcd/m_{c,b})$ corrections are known for the SM currents and
would be straightforward to calculate for any new physics, but are
neglected below.

The three heavy quark spin symmetry doublets relevant for this paper can be
represented by the (super)fields, which have the correct transformation
properties under Lorentz and heavy quark symmetries~\cite{Falk:1991nq},
\begin{align}
H_v & = \frac{1 + \slashed{v}}2 \Big[ B^*_v\, \slashed{\epsilon}
  - B_v\, \g^5 \Big] , \nn\\*
K_v & = \frac{1 + \slashed{v}}2 \Big[ V_v \g^5 \slashed{\epsilon}
  + P_v \Big] , \\
F^\alpha_v & = \frac{1 + \slashed{v}}2 
  \bigg\{ T_v\, \epsilon^{\alpha \beta} \g_\beta
  - V_v\, \sqrt{\frac32}\, \g^5 \bigg[ \epsilon^\alpha
  - \frac13 \slashed{\epsilon}(\g^\alpha - v^\alpha) \bigg] \bigg\} . \nn
\end{align}
In this paper each representation occurs for only one heavy quark flavor, so
for simplicity we denote the components of $H_v$ by $B_v$ and $B^*_v$.
The $\epsilon^{\alpha\beta}$ denote a normalized traceless symmetric spin-2
polarization tensor. 

Similar to Ref.~\cite{Falk:1992wt}, including $\lqcd/m_{c,b}$
corrections, the $\Bbar \to \dss$ matrix elements can be written as
\begin{widetext}
\begin{subequations}\label{eqn:TraceLM}
\begin{align}
\frac{\langle \dsW |\, \cbar\, \Gamma\, b\, | \Bbar \rangle}{\sqrt{m_{\dsW} m_B}}
& = \zeta(w)\, \bigg\{ \text{Tr} \big[ \bar K_{v'}\, \Gamma\, H_v \big]
  + \varepsilon_c\, \text{Tr}\big[ \bar K^{(1)}_{v',v}\, \Gamma\, H_v \big]
  - \varepsilon_b\, \hat G_b\, \text{Tr} \Big[ \bar K_{v'} \Gamma\,
  \frac{1-\slashed{v}}2\, \g^5 B_v \Big]\bigg\}\,, \label{eqn:TraceW}\\
\frac{\langle \dsN |\, \cbar\, \Gamma\, b\, | \Bbar \rangle}{\sqrt{m_{\dsN} m_B}}
& = \tau(w)\, \bigg\{ \text{Tr} \big[ v_\sigma \bar F^{\sigma}_{v'}\, \Gamma\, H_v \big]
  + \varepsilon_c\, \text{Tr}\big[ \bar F^{(1)}_{v',v}\, \Gamma\, H_v \big]
  + \varepsilon_b\, \hat F_b\, \text{Tr} \Big[ v_\sigma \bar F^{\sigma}_{v'}\, 
  \Gamma\, \frac{1-\slashed{v}}2\, \g^5 B_v \Big]\bigg\}\,, \label{eqn:TraceN}
\end{align}
\end{subequations}
where $\varepsilon_{c,b} = 1/(2m_{c,b})$, $\Gamma$ is an arbitrary Dirac
matrix, and
\begin{subequations}\label{MNdef}
\begin{align}
\bar{K}^{(1)}_{v',v} & = \Big[ V_{v'} \g^5 \big( \slashed{\epsilon} \hat M_2 + \epsilon \cdot v\, \hat M_3 \big) + P_{v'} \hat M_1 \Big] \frac{1 + \slashed{v}'}{2}
  + \Big[ V_{v'} \g^5 \big( \slashed{\epsilon} \hat M_5 + \epsilon \cdot v\, \hat M_6 \big) + P_{v'} \hat M_4 \Big] \frac{1 - \slashed{v}'}{2}\,, \\
\bar{F}^{(1)}_{v',v} & = \Big[ T_{v'} \big( \epsilon^{\mu\nu}\g_\mu \g_\nu\, \hat N_1 + \epsilon^{\mu\nu} v_\mu \g_\nu\, \hat N_2 + \epsilon^{\mu\nu} v_\mu v_\nu\, \hat N_3 \big)
  + \frac{V_{v'}}{\sqrt{6}}\, \big( \slashed{\epsilon} \hat N_4 + \epsilon \cdot v\, \hat N_5 \big) \g^5 \Big] \frac{1 + \slashed{v}'}{2} \nn\\*
& + \Big[ T_{v'} \big( \epsilon^{\mu\nu}\g_\mu \g_\nu\, \hat N_6 + \epsilon^{\mu\nu} v_\mu \g_\nu\, \hat N_7 + \epsilon^{\mu\nu} v_\mu v_\nu\, \hat N_{8} \big)
  + \frac{V_{v'}}{\sqrt{6}}\, \big( \slashed{\epsilon} \hat N_9 + \epsilon \cdot v\, \hat N_{10} \big) \g^5 \Big] \frac{1-\slashed{v}'}{2}\,.
\end{align}
\end{subequations}
\end{widetext}

At leading order, in the heavy quark limit, all $\Bbar \to \dsW$ form factors are
determined by one Isgur-Wise function, $\zeta(w)$, while all $\Bbar \to \dsN$ form
factors are determined by another, $\tau(w)$.  (In the notation of Ref.~\cite{Isgur:1990jf}, $\zeta(w)$ is twice the
function $\tau_{1/2}$ and $\tau(w)$ is $\sqrt3$ times the function
$\tau_{3/2}$.)  All form factors are viewed as
functions of the dimensionless kinematic variable $w$, instead of $q^2 = (p_B
-p_{\dss})^2$, with
\begin{equation}\label{wdef}
	w = v\cdot v' = \frac{m_B^2  + m_{\dss}^2 - q^2}{2m_B m_{\dss}}\,.
\end{equation}
Here $v = p_B / m_B$ and $v' = p_{\dss} / m_{\dss}$ are the four-velocities
of the initial and final states.

The coefficients $\hat M_i$ and $\hat N_i$ contain order $\lqcd/m_{c,b}$
corrections. These are expressed in terms of subleading Isgur-Wise functions,
which arise either from corrections to the HQET Lagrangian or from matching the
current operators onto HQET~\cite{Eichten:1990vp, Luke:1990eg, Falk:1990pz}. 
Specifically: (i)~matrix elements of the ${\cal O}(\lqcd/m_{c,b})$ current
operators give rise to the subleading (dimensionful) Isgur-Wise functions
$\zeta_1$ and $\tau_{1,2}$; (ii) matrix elements involving the $\lqcd/m_{c,b}$
suppressed kinetic energy operator, $\bar h_v (i D)^2 h_v/(2m_Q)$, are spin
symmetry conserving, and generate the functions $\chi_{\rm ke}^{c,b}$ and
$\eta_{\rm ke}^{c,b}$; (iii) matrix elements involving the chromomagnetic
operator in the HQET Lagrangian, $(g_s/2)\,\bar h_v \sigma_{\mu\nu} G^{\mu\nu}
h_v / (2m_Q)$, which violates spin symmetry, generate the functions
$\chi_{1,2}^{c,b}$ and $\eta_{1,2,3}^{c,b}$. The notation for these subleading
Isgur-Wise functions are summarized in Table~\ref{IWfnlist}.

\begin{table}[b]
\renewcommand*{\arraystretch}{1.2}
\begin{tabular}{c|c|cc}
\hline\hline
doublet  &  leading order  &  $1/m$ current  &  $1/m$ Lagrangian  \\
\hline 
$\dsW$  &  $\zeta$  &  $\zeta_{1}$  &  $\chi_{\rm ke}^{c,b}$, 
  $\chi_{1,2}^{c,b}$ $\to$ $\chi_{1,2}$\\
$\dsN$  &  $\tau$  &  $\tau_{1,2}$  &  $\eta_{\rm ke}^{c,b}$, 
  $\eta_{1,2,3}^{c,b}$ $\to$ $\eta_{1,2,3}$\\
\hline\hline
\end{tabular}
\caption{Leading and subleading Isgur-Wise functions that parametrize $\Bbar \to
\dss$ form factors at $\mathcal{O}(\lqcd/m_{c,b})$.  The arrows in the last
column indicate the minimal set of functions needed for the $1/m$ Lagrangian
corrections, if the replacements in Eq.~\eqref{IWreplace} are made.  Omitted
upper indices mean~`$c$'.}
\label{IWfnlist}
\end{table}

As we consider only $B$ (and not $B^*$) decays, the $\lqcd/m_b$ corrections from
the chromomagnetic operator in the  Lagrangian --- the terms involving
$\chi_{1,2}^{b}$ ($\eta_{1,2,3}^{b}$) --- enter in just one linear
combination for all $\dsW$ ($\dsN$) form factors, as do the heavy quark spin 
symmetry conserving subleading Isgur-Wise functions, $\chi_{\rm ke}^{c,b}$
($\eta_{\rm ke}^{c,b}$). These can therefore be absorbed into the leading
order Isgur-Wise functions via the replacements,
\begin{align}\label{IWreplace}
&\zeta + \varepsilon_c\, \chi_{\rm ke}^{c}
  + \varepsilon_b\big[\chi_{\rm ke}^{b} + 6\chi_1^{b}
  - 2(w+1)\chi_2^{b}\big] \to \zeta\,, \\*
& \tau + \varepsilon_c\, \eta_{\rm ke}^{c}
  + \varepsilon_b\big[\eta_{\rm ke}^{b} + 6 \eta_1^{b}
  - 2(w-1)\eta_2^{b} + \eta_3^{b}\big] \to \tau\,. \nn
\end{align}
These replacements formally introduce ${\cal O}(\lqcd^2/m_{c,b}^2)$ errors.
Because the $\lqcd/m_{c,b}$ terms themselves can be dominant near zero recoil,
these ${\cal O}(\lqcd^2/m_{c,b}^2)$ corrections may in practice be sizable. 

Hereafter it is understood that the replacements in Eq.~\eqref{IWreplace}
are made, unless explicitly noted otherwise. As in Table~\ref{IWfnlist}, we
omit the $c$ superscript from the remaining (unabsorbed) subleading
Isgur-Wise functions.  We further define $\hat\chi_{1,2} =
\chi_{1,2}/\zeta$, $\hat\zeta_{1} = \zeta_{1}/\zeta$, $\hat\eta_{1,2,3} =
\eta_{1,2,3}/\tau$, and $\hat\tau_{1,2} = \tau_{1,2}/\tau$. With these
conventions, the  $\hat M_i$ and $\hat N_i$ coefficient functions in
Eq.~(\ref{MNdef}) are
\begin{align}
\hat M_1 & = 6 \hat\chi_1 - 2\hat\chi_2(w+1)\,, \qquad
  \hat M_2 = -2 \hat\chi_1\,, \nn\\*
\hat M_3 &= 2\hat\chi_2\,,\qquad
  \hat M_4 = 2\hat\zeta_1(w+1) - 3\, \frac{\bar\Lambda^*w - \bar\Lambda}{w+1} \,, 
  \nn\\*
\hat M_{5} &= -\frac{\bar\Lambda^*w - \bar\Lambda}{w+1}\,, \qquad 
  \hat M_6 = -2\hat\zeta_1\,, \nn\\*
\hat G_b & = \frac{(1+2w)\bar\Lambda^*-(2+w)\bar\Lambda}{w+1} - 2(w-1)\,\hat\zeta_1(w) \,,
\end{align}
and
\begin{align}
\hat N_1 & = -\hat\eta_3(w+1)\,,\qquad 
  \hat N_2 = \hat\eta_3 - 2\hat\eta_1\,,\nn\\*
\hat N_3 &= -2 \hat\eta_2\,, \qquad
  \hat N_4 = -(2\hat\eta_1 + 3\hat\eta_3)(w+1)\,,\nn\\*
\hat N_5 &= 10\hat\eta_1 + 4\hat\eta_2(w-1) -5\hat\eta_3\,, \nn\\*
\hat N_6 &= (\hat\tau_1-\hat\tau_2)(w-1) - (\bar\Lambda'w - \bar\Lambda)\,,\nn\\*
\hat N_7 &= -(\hat\tau_1 -\hat\tau_2)\,,\qquad \hat N_8 = -2 \hat\tau_1\,, \nn \\*
\hat N_9 &= 3(\hat\tau_1 -\hat\tau_2)(w-1) - 4(\bar\Lambda'w -\bar\Lambda)\,,\nn\\*
\hat N_{10} &= \hat\tau_1(4w-1) + 5\hat\tau_2 \,, \nn\\*
\hat F_b &= \bar\Lambda + \bar\Lambda' - (2w+1) \hat\tau_1 - \hat\tau_2\,.
\end{align}
The $\lqcd/m_b$ corrections not absorbed into the leading order
Isgur-Wise functions occur exclusively in the $\hat F_b$ and $\hat G_b$ linear
combinations.  (The sign difference between these terms in
Eqs.~(\ref{eqn:TraceW}) and (\ref{eqn:TraceN}) is simply due to defining
their known $\bar\Lambda^{(\prime, *)}$ parts to be positive.)

\subsection{\texorpdfstring{$\Bbar \to \dsW$}{BW} form factors}

We define form factors in agreement with those in
Refs.~\cite{Leibovich:1997tu, Leibovich:1997em} for the SM currents. 
For $\Bbar \to \dSs$
\begin{align}
\ampB{\dSs}{\cbar\,b} &= \ampB{\dSs}{\cbar \g_\mu b} = 0\,, \nn\\
\ampB{\dSs}{\cbar\g_5 b} &= \sqrt{m_{\dSs} m_B}\, g_P\,, \nn\\
\ampB{\dSs}{\cbar \g_\mu \g_5 b} &= \sqrt{m_{\dSs} m_B} 
  \big[ g_+ (v_\mu+v'_\mu) + g_- (v_\mu-v'_\mu) \big] , \nn\\
\ampB{\dSs}{\cbar \sigma_{\mu\nu} b} &= 
  \sqrt{m_{\dSs} m_B}\, g_T\, \varepsilon_{\mu\nu\alpha\beta}\, 
  v^\alpha v'^\beta \,,
\end{align}
and for $\Bbar \to \dVs$,
\begin{align}\label{formf121}
\ampB{\dVs}{\cbar\,b} &= -\sqrt{m_{\dVs} m_B}\, g_S\, (\epsilon^*\cdot v)\,, \nn\\
\ampB{\dVs}{\cbar\g_5 b} &= 0\,, \nn\\
\ampB{\dVs}{\cbar \g_\mu b} &=  \sqrt{m_{\dVs} m_B}\, 
  \big[ g_{V_1}\, \epsilon^*_\mu \nn\\*
  & \qquad + (g_{V_2} v_\mu + g_{V_3} v'_\mu)\, (\epsilon^* \cdot v) \big]\,, \nn\\
\ampB{\dVs}{\cbar \g_\mu \g_5 b} &= i \sqrt{m_{\dVs} m_B}\, g_A\, 
  \varepsilon_{\mu\alpha\beta\gamma}\, \epsilon^{*\alpha} v^\beta\, v'^\gamma
  \,, \nn\\
\ampB{\dVs}{\cbar \sigma_{\mu\nu} b} &= i \sqrt{m_{\dVs} m_B}\,
  \big[ g_{T_1} (\epsilon^*_\mu v_\nu - \epsilon^*_\nu v_\mu) \nn\\*
  & \qquad + g_{T_2} (\epsilon^*_\mu v'_\nu - \epsilon^*_\nu v'_\mu) \nn\\*
  & \qquad + g_{T_3} (\epsilon^*\cdot v) (v_\mu v'_\nu - v_\nu v'_\mu) \big]\,.
\end{align}
These form factors are dimensionless functions of $w$, and hereafter we often
suppress displaying the $w$ variable.  In the heavy quark limit, each of
these form factors either vanishes or is determined by the Isgur-Wise
function,
\begin{align}\label{BD12lead}
g_+ &= g_{V_2} = g_{T_3} = 0 \,, \qquad
  g_P = g_{V_1} = (w-1)\, \zeta\,, \nn\\*
g_- &= g_T = g_S = g_A = -g_{V_3} = -g_{T_1} = g_{T_2} = \zeta\,.
\end{align}
Unlike the leading order Isgur-Wise function in $\Bbar \to D^{(*)}$ decays, the
function $\zeta(w)$ is not subject to any symmetry imposed normalization
condition. Near zero recoil, only $g_P$, $g_{V_1}$, and the linear combination
$g_{T_1}+g_{T_2}$ contribute to the decay rates without a $(w-1)$ suppression
(in addition to the $\sqrt{w^2-1}$ phase space factor; see 
Eqs.~\eqref{eqn:D0srate} and~\eqref{eqn:D1srate} below), so heavy quark symmetry
implies that these form factors in the heavy quark limit are suppressed near
zero recoil as $(w-1)\, \zeta$.  This is why the order $\lqcd/m_{c,b}$
terms have enhanced significance for these decays~\cite{Leibovich:1997tu,
Leibovich:1997em}.  At order $\lqcd^2/m_{c,b}^2$ and higher, the expansion is
expected to behave as suggested by the power counting.

To write the ${\cal O}(\lqcd/m_{c,b})$ corrections in a compact form, 
we define
\beq\label{gihat}
  \hat g_i(w) = g_i(w) / \zeta(w) \,.
\eeq
Denoting $\haS = \aS/\pi$ we obtain
\begin{widetext}
\begin{align}\label{BD0m}
\hat g_P &= (w-1) \big( 1 + \haS C_P \big)
  + \varepsilon_c \big\{ 3(w \bar\Lambda^* - \bar\Lambda) - 2(w^2-1)\hat\zeta_1 
  + (w-1)[6\hat\chi_1-2(w+1)\hat\chi_2] \big\}
  - \varepsilon_b\, (w+1)\, \hat G_b\,, \nn\\*
\hat g_+ &= \haS\, (w-1)\, \frac{C_{A_2}+C_{A_3}}2
  - \varepsilon_c \bigg[ 3\frac{w\bar\Lambda^*-\bar\Lambda}{w+1}
  - 2(w-1)\hat\zeta_1 \bigg] - \varepsilon_b\, \hat G_b \,, \nn\\*
\hat g_- &= 1 + \haS \bigg[ C_{A_1} + (w-1)\frac{C_{A_2}-C_{A_3}}2 \bigg]
  + \varepsilon_c [6\hat\chi_1-2(w+1)\hat\chi_2] \,, \nn\\*
\hat g_T &= 1 + \haS\, C_{T_1}
  + \varepsilon_c \bigg[ 3\frac{w\bar\Lambda^*-\bar\Lambda}{w+1} 
  -2 (w-1)\hat\zeta_1 + 6\hat\chi_1-2(w+1)\hat\chi_2 \bigg] 
  - \varepsilon_b\, \hat G_b \,.
\end{align}
The $C_i$'s encode the $\aS$ corrections (and are given in Appendix~A in
Ref.~\cite{Bernlochner:2017jka}), and correspond to integrating out the $b$ and
$c$ quarks at a common scale, chosen as $\mu = \sqrt{m_c m_b}$. For the order 
$\lqcd/m_{c,b}$ and $\alpha_s$ contributions to the $\Bbar\to\dVs$ form
factors we obtain
\begin{align}
\label{BD1sm}
\hat g_S &= 1 + \haS\, C_S
  - \varepsilon_c \bigg[ \frac{w \bar\Lambda^*-\bar\Lambda}{w+1} 
  - 2(w-1) \hat\zeta_1 + 2\hat\chi_1 - 2(w+1)\hat\chi_2 \bigg]
  - \varepsilon_b\, \hat G_b\,, \nn\\*
\hat g_{V_1} &= (w-1) \big( 1 + \haS\, C_{V_1} \big)
  + \varepsilon_c \big[ w \bar\Lambda^*-\bar\Lambda - 2(w-1)\hat\chi_1 \big]
  - \varepsilon_b\, (w+1)\, \hat G_b \,,\nn\\
\hat g_{V_2} &= - \haS\, C_{V_2}
  + \varepsilon_c \big( 2\hat\zeta_1 - 2\hat\chi_2 \big) \,,\nn\\
\hat g_{V_3} &= - 1 - \haS\, (C_{V_1} + C_{V_3})
  - \varepsilon_c \bigg( \frac{w \bar\Lambda^*-\bar\Lambda}{w+1} 
  + 2 \hat\zeta_1 - 2\hat\chi_1 + 2\hat\chi_2 \bigg)
  + \varepsilon_b\, \hat G_b \,,\nn\\
\hat g_A &= 1 + \haS\, C_{A_1}
  + \varepsilon_c \bigg( \frac{w\bar\Lambda^*-\bar\Lambda}{w+1}
  - 2\hat\chi_1 \bigg) - \varepsilon_b\, \hat G_b\,, \nn\\
\hat g_{T_1} &= - 1 - \haS \big[ C_{T_1} + (w-1)\, C_{T_2} \big]
  + \varepsilon_c \bigg( \frac{w \bar\Lambda^*-\bar\Lambda}{w+1} 
  + 2\hat\chi_1 \bigg) + \varepsilon_b\, \hat G_b\,,\nn\\
\hat g_{T_2} &= 1 + \haS \big[ C_{T_1} - (w-1)\, C_{T_3} \big]
  + \varepsilon_c \bigg( \frac{w \bar\Lambda^*-\bar\Lambda}{w+1} 
  - 2\hat\chi_1 \bigg) + \varepsilon_b\, \hat G_b \,,\nn\\*
\hat g_{T_3} &= - \haS\, C_{T_2}
  + \varepsilon_c \big( 2\hat\zeta_1 + 2\hat\chi_2 \big)  \,.
\end{align}
\end{widetext}
In Eqs.~\eqref{BD0m} and~\eqref{BD1sm} the expressions for the SM terms ($g_+$,
$g_-$, $g_{V_1}$, $g_{V_2}$, $g_{V_3}$, and $g_A$) agree with
Refs.~\cite{Leibovich:1997tu, Leibovich:1997em}.  Only four functions of $w$ are
needed to parametrize all twelve $\Bbar \to \dsW$ form factors in Eqs.~(\ref{BD0m}) and
\eqref{BD1sm} at this order: $\zeta$, $\hat\zeta_1$, and $\hat\chi_{1,2}$.  Only
the $6\hat\chi_1 - 2(w+1)\hat\chi_2 = \hat M_1$ linear combination of
$\hat\chi_1$ and $\hat\chi_2$ occurs for $\Bbar \to D_0^*$ in Eq.~\eqref{BD0m}, as
expected from Eqs.~\eqref{MNdef}.

\subsection{\texorpdfstring{$\Bbar \to \dsN$}{BN} form factors}

We define the form factors for $\Bbar \to \dsN$ such that they agree with those in
Refs.~\cite{Leibovich:1997tu, Leibovich:1997em} for the SM terms,
\begin{align}\label{formf321}
\ampB{\dV}{\cbar\,b} &= \sqrt{m_{\dV} m_B}\, f_S\, (\epsilon^*\cdot v) \,,\nn\\
\ampB{\dV}{\cbar\g_5 b} &= 0\,, \nn\\
\ampB{\dV}{\cbar\g_\mu b} &= \sqrt{m_{\dV} m_B}\, \big[ f_{V_1}\, \epsilon^*_\mu\nn\\* 
  &\qquad + (f_{V_2} v_\mu + f_{V_3} v'_\mu) (\epsilon^*\cdot v) \big] , \nn\\
\ampB{\dV}{\cbar\g_\mu\g_5 b} &= i\, \sqrt{m_{\dV} m_B}\, f_A\,
\varepsilon_{\mu\alpha\beta\gamma} \epsilon^{*\alpha} v^\beta v'^\gamma \,,
\nn\\
\ampB{\dV}{\cbar \sigma_{\mu\nu} b} &= i \sqrt{m_{\dV} m_B} \big[ f_{T_1}
(\epsilon^*_\mu v_\nu - \epsilon^*_\nu v_\mu) \nn\\*
  &\qquad + f_{T_2} (\epsilon^*_\mu v'_\nu - \epsilon^*_\nu v'_\mu) \nn\\*
  &\qquad + f_{T_3} (\epsilon^*\cdot v) (v_\mu v'_\nu - v_\nu v'_\mu) \big],
\end{align}
while for  $\Bbar \to \dTs$,
\begin{align}
\ampB{\dTs}{\cbar\,b} &= 0 \,, \nn\\
\ampB{\dTs}{\cbar\g_5 b} &= \sqrt{m_{\dTs} m_B}\,
  k_P\,\epsilon^*_{\alpha\beta}\, v^\alpha v^\beta \,, \nn\\
\ampB{\dTs}{\cbar\g_\mu b} &= i \sqrt{m_{\dTs} m_B}\, k_V\,
  \varepsilon_{\mu\alpha\beta\gamma}\,\epsilon^{*\alpha\sigma} v_\sigma v^\beta
  v'^\gamma \,,\nn\\
\ampB{\dTs}{\cbar\g_\mu\g_5 b}  &= \sqrt{m_{\dTs} m_B}\, \big[ k_{A_1}\,
  \epsilon^*_{\mu\alpha} v^\alpha \nn\\*
  &\qquad + (k_{A_2} v_\mu + k_{A_3} v'_\mu)\,
  \epsilon^*_{\alpha\beta}\, v^\alpha v^\beta \big] , \nn\\
\ampB{\dTs}{\cbar \sigma_{\mu\nu} b}  &= \sqrt{m_{\dTs} m_B}\,
  \varepsilon_{\mu\nu\alpha\beta}\, \big\{ [k_{T_1} (v+v')^\alpha \nn\\
  &\qquad + k_{T_2} (v-v')^\alpha)]\, \epsilon^{*\gamma\beta} v_\gamma \nn\\
  &\qquad + k_{T_3}\, v^\alpha v'^\beta \epsilon^{*\rho\sigma} v_\rho v_\sigma\big\}\,.\label{formf322}
\end{align}
The form factors $f_i$ and $k_i$ are again dimensionless functions of $w$. In
the heavy quark limit, each of these form factors either vanishes or is
determined by the Isgur-Wise function, $\tau(w)$. (This Isgur-Wise function
is different from $\zeta(w)$, although model calculations can relate the
two).  The simple parametrizations in Eqs.~\eqref{formf321}
and~\eqref{formf322} yield the slightly complicated relations
\begin{align}\label{BD32lead}
k_{A_2} &= k_{T_2} = k_{T_3} = 0\,,\qquad
  f_{V_1} = (1-w^2)\tau/\sqrt6 \,, \nn\\*
f_{V_3} &= (w-2)\,\tau/\sqrt6\,, \qquad
  -f_{V_2} = f_{T_3} = 3\tau/\sqrt6\,, \nn\\*
f_A & = f_S/2 = -f_{T_1} = f_{T_2} = k_{A_1}/\sqrt6 = -(w+1)\,\tau/\sqrt6\,,\nn\\
k_P & = -k_V = k_{A_3} = k_{T_1} = \tau\,.
\end{align}
At zero recoil where $w=1$ and $v=v'$, only the $f_{V_1}$ form factor can
contribute (as well as the linear combination $f_{T_1}+f_{T_2}$), since
$\epsilon^\mu v'_\mu$ and $\epsilon^{\mu\nu}v'_\nu$ vanish.  Heavy quark
symmetry implies that $f_{V_1}(1)$ is either of order $\lqcd/m_{c,b}$, or its
dependence on the leading Isgur-Wise function, $\tau(w)$, is suppressed by
$(w-1)$~\cite{Isgur:1990jf}.  This is why, as explained above, the ${\cal
O}(\lqcd/m_{c,b})$ terms are so significant for semileptonic decays to
excited charmed mesons.

We define in analogy with Eq.~\eqref{gihat},
\begin{equation}
\hat f_i(w) = f_i(w) / \tau(w) \,, \qquad 
  \hat k_i(w) = k_i(w) / \tau(w) \,.
\end{equation}
For the order $\lqcd/m_{c,b}$ and $\alpha_s$ contributions to the $\Bbar \to \dV$ form factors, we obtain
\begin{widetext}
\begin{align}\label{BD1m}
\sqrt6\, \hat f_S &= -2(w+1) (1 + \haS C_S)
  - \varepsilon_b\, 2(w-1) \hat F_b \nn\\*
  & \quad - \varepsilon_c \big\{ 4(w \bar\Lambda'-\bar\Lambda)
  - 2(w-1) [(2w+1)\hat\tau_1 + \hat\tau_2] 
  + 2(w+1) [6\hat\eta_1 + 2(w-1)\hat\eta_2 - \hat\eta_3] \big\} , \nn\\*
\sqrt6\, \hat f_{V_1} &= (1-w^2)\, (1 + \haS\, C_{V_1})
  - \varepsilon_b\, (w^2-1) \hat F_b
  - \varepsilon_c \big[ 4(w+1)(w \bar\Lambda'-\bar\Lambda)
  - (w^2-1)(3\hat\tau_1-3\hat\tau_2+2\hat\eta_1+3\hat\eta_3) \big] , \nn\\*
\sqrt6\, \hat f_{V_2} &= - 3 - \haS
  \big[3\, C_{V_1}+ 2(1+w)\, C_{V_2} \big] - \varepsilon_b\, 3 \hat F_b
  - \varepsilon_c \big[ (4w-1)\hat\tau_1 + 5\hat\tau_2 + 10\hat\eta_1
    + 4(w-1)\hat\eta_2-5\hat\eta_3 \big] , \nn\\*
\sqrt6\, \hat f_{V_3} &= w-2 - \haS
  \big[ (2-w)\, C_{V_1} + 2(1+w)\, C_{V_3}\big]
  + \varepsilon_b\, (2+w) \hat F_b \nn\\*
  & \quad + \varepsilon_c \big[ 4(w \bar\Lambda'-\bar\Lambda) +
  (2+w)\hat\tau_1 + (2+3w)\hat\tau_2 - 2(6+w)\hat\eta_1 - 4(w-1)\hat\eta_2 
  - (3w-2)\hat\eta_3 \big] , \nn\\*
\sqrt6\, \hat f_A &= -(w+1) (1 + \haS\, C_{A_1})
  - \varepsilon_b\, (w-1) \hat F_b
  - \varepsilon_c \big[ 4(w\bar\Lambda'-\bar\Lambda)
  - 3(w-1)(\hat\tau_1-\hat\tau_2) - (w+1)(2\hat\eta_1+3\hat\eta_3) \big] ,\nn\\
\sqrt6\, \hat f_{T_1} &= (w+1) \bigg[1 + \haS 
  \big[ C_{T_1} + (w-1)\, C_{T_2} \big] \bigg]
  + \varepsilon_b\, (w-1) \hat F_b \nn\\*
  & \quad - \varepsilon_c \big[ 4(w \bar\Lambda'-\bar\Lambda)
  - 3(w-1) (\hat\tau_1-\hat\tau_2) + (w+1)(2\hat\eta_1+3\hat\eta_3) \big] ,\nn\\
\sqrt6\, \hat f_{T_2} &= -(w+1) \bigg[1 + \haS
  \big[ C_{T_1} - (w-1)\, C_{T_3} \big] \bigg]
  + \varepsilon_b\, (w-1) \hat F_b \nn\\*
  & \quad - \varepsilon_c \big[ 4(w \bar\Lambda'-\bar\Lambda)
  - 3(w-1) (\hat\tau_1-\hat\tau_2) - (w+1)(2\hat\eta_1+3\hat\eta_3) \big] ,\nn\\
\sqrt6\, \hat f_{T_3} &= 3 + \haS 
  \big[ 3\, C_{T_1} - (2-w)\, C_{T_2} + 3\, C_{T_3} \big]
  + \varepsilon_b\, 3 \hat F_b 
  - \varepsilon_c \big[ (4w-1)\hat\tau_1 + 5 \hat\tau_2 
  - 10\hat\eta_1 - 4(w-1)\hat\eta_2 + 5\hat\eta_3  \big]\,.
\end{align}
\end{widetext}
For the order $\lqcd/m_{c,b}$ and $\alpha_s$ contributions to the
$\Bbar \to\dTs$ form factors we obtain
\begin{align}\label{BD2m}
\hat k_P &= 1 + \haS\, C_P
  + \varepsilon_b\, \hat F_b \nn\\*
  &\quad + \varepsilon_c \big[ (2w+1)\hat\tau_1 + \hat\tau_2 
    -2\hat\eta_1 -2(w-1)\hat\eta_2 + \hat\eta_3 \big] , \nn\\
\hat k_V &= - 1 - \haS\, C_{V_1}
  - \varepsilon_b\, \hat F_b - \varepsilon_c \big( \hat\tau_1 - \hat\tau_2 
  - 2\hat\eta_1 + \hat\eta_3 \big) , \nn\\
\hat k_{A_1} &= -(w+1) \big(1 + \haS\, C_{A_1} \big)
  - \varepsilon_b (w-1) \hat F_b \nn\\*
  &\quad - \varepsilon_c \big[ (w-1)(\hat\tau_1-\hat\tau_2) 
  - (w+1)(2\hat\eta_1-\hat\eta_3) \big] , \nn\\*
\hat k_{A_2} &= \haS\, C_{A_2}
  - \varepsilon_c\, 2 \big( \hat\tau_1 + \hat\eta_2 \big) , \nn\\
\hat k_{A_3} &= 1 + \haS\, (C_{A_1} + C_{A_3})
  + \varepsilon_b\, \hat F_b \nn\\*
  &\quad - \varepsilon_c \big( \hat\tau_1 + \hat\tau_2 
  + 2 \hat\eta_1 - 2 \hat\eta_2 - \hat\eta_3 \big) , \nn\\*
\hat k_{T_1} &= 1 + \haS \bigg[ C_{T_1} 
  + \frac{w-1}2\, (C_{T_2}-C_{T_3}) \bigg]
  - \varepsilon_c \big( 2\hat\eta_1 - \hat\eta_3 \big) ,\nn\\
\hat k_{T_2} &= \haS\, \frac{w+1}2\, (C_{T_2}+C_{T_3})
  + \varepsilon_b\, \hat F_b - \varepsilon_c \big(\hat\tau_1-\hat\tau_2 \big) ,\nn\\
\hat k_{T_3} &= - \haS C_{T_2}
  + \varepsilon_c\, 2 \big(\hat\tau_1-\hat\eta_2 \big) .
\end{align}
In Eqs.~(\ref{BD1m}) and (\ref{BD2m}) the relations for the SM terms
($f_{V_1}$, $f_{V_2}$, $f_{V_3}$, $f_A$, $k_{V_1}$, $k_{V_2}$, $k_{V_3}$, and
$k_A$) agree with Refs.~\cite{Leibovich:1997tu, Leibovich:1997em}. In this
case six functions of $w$ are needed to parametrize all sixteen $\Bbar \to \dsN$ form
factors in Eqs.~(\ref{BD1m}) and (\ref{BD2m}), including all ${\cal
O}(\lqcd/m_{c,b})$ corrections: $\tau$, $\hat\tau_{1,2}$, and
$\hat\eta_{1,2,3}$.

\subsection{Equations of motion}

As in Ref.~\cite{Bernlochner:2017jka}, we can verify the relations stemming
from the QCD equations of motion between the (pseudo)scalar
and the (axial)vector matrix elements,
\begin{align}\label{currentrel}
-[\ov m_b(\mu)+\ov m_c(\mu)]\, \ampBs{\dSs}{\cbar \g_5 b} 
  	&=  \ampBs{\dSs}{\cbar \slashed{q} \g_5 b} \,, \nn\\*
[\ov m_b(\mu)-\ov m_c(\mu)]\, \ampBs{\dVs}{\cbar b}  
	&=  \ampBs{\dVs}{\cbar \slashed{q} b} \,, \nn\\*
-[\ov m_b(\mu)+\ov m_c(\mu)]\, \ampBs{\dTs}{\cbar \g_5 b} 
	&=  \ampBs{\dTs}{\cbar \slashed{q} \g_5 b}\,, \nn\\*
[\ov m_b(\mu)-\ov m_c(\mu)]\, \ampBs{\dV}{\cbar b} 
	&= \ampBs{\dV}{\cbar \slashed{q} b}\,.
\end{align}
Using $m_b = m_B - \bar\Lambda + {\cal O}(\lqcd^2/m_b)$ and $m_c = m_{\dss}
- \bar\Lambda^{\prime,*} + {\cal O}(\lqcd^2/m_c)$ imply that all first order
$\lqcd/m_{c,b}$ and $\aS$ corrections agree in Eq.~\eqref{currentrel} as
they must.  Note that the left-hand sides of Eq.~\eqref{currentrel} contain the
running quark masses at the common scale $\mu$.  At order $\aS$ the
results are sensitive to this, and the ${\cal O}(\aS)$ terms from the
expansion of
\begin{equation}
m_Q = \ov m_Q(\mu) \bigg[ 1+ \frac{\aS(\mu)}\pi\,
  \bigg( \frac43 - \ln \frac{m_Q^2}{\mu^2}\bigg) + \ldots \bigg] \,,
\end{equation}
are required to be present for Eq.~(\ref{currentrel}) to be satisfied. As
emphasized in Ref.~\cite{Bernlochner:2017jka}, it is probably better to
evaluate the scalar and pseudoscalar matrix elements using Eqs.~\eqref{BD0m},
\eqref{BD1sm}, \eqref{BD1m}, and \eqref{BD2m} instead of Eq.~\eqref{currentrel},
because the natural choice for $\mu$ is below $m_b$.  In the $\ov{\rm MS}$
scheme fermions do not decouple for $\mu < m$, introducing artificially large
corrections in the running, compensated by corresponding spurious terms in
the $\beta$-function computed without integrating out heavy
quarks~\cite{Manohar:1996cq}.

\section{\texorpdfstring{$B\to \dss\ell\bar\nu$}{BDss} rates for generic NP}
\label{sec:nprates}

It is straightforward to calculate the $B\to \dss\ell\bar\nu$ rates including
lepton mass effects and all possible four-fermion operators. The double
differential distributions in the SM were written down in
Ref.~\cite{Bernlochner:2016bci}, the integrals of which agree with the
expressions below.  Here we give the single differential distributions as the SM
plus generic NP.  These expressions can be used with any form factor input to
study the SM predictions and possible patterns of NP in $R(\dss)$. In
Appendix~\ref{app:NPA} we provide explicit results for the $\Bbar \to D^{**}
\ell \bar\nu$ amplitudes, which may be used in combination with $\dss$ and
$\tau$ decay amplitudes to derive fully differential distributions of the
visible decay products in $\Bbar \to D^{**}\ell\bar\nu$, including all
interference effects~\cite{Ligeti:2016npd}. 

We consider the following complete basis for the four-Fermi operators
mediating $b \to c \ell \bar\nu$ decay
\begin{subequations}\label{abtdef}
\begin{align}
\text{SM:~~} & i2\sqrt{2}\, V_{cb} G_F\big[\cbar \g^\mu P_L b\big] \big[\bar\ell \g_\mu P_L \nu\big]\,, \\*
\text{Vector:~~} & i2\sqrt{2}\, V_{cb} G_F
  \big[\cbar\big(\alVLt \g^\mu P_L + \alVRt \g^\mu P_R\big)b\big] \nn\\*
&\qquad\quad \times
  \big[\bar\ell\big(\beVLt \g_\mu P_L + \beVRt \g_\mu P_R\big) \nu\big]\,, \\
\text{Scalar:~~} & i2\sqrt{2}\, V_{cb} G_F
  \big[\cbar\big(\alSLt P_L + \alSRt P_R\big)b\big] \nn\\*
&\qquad\quad \times
  \big[\bar\ell\big(\beSLt P_L + \beSRt P_R\big) \nu\big]\,, \\
\text{Tensor:~~} & i2\sqrt{2}\, V_{cb} G_F
  \big[ \big(\cbar\, \alTLt \sigma^\mn P_L b\big)
  \big(\bar\ell\, \beTLt \sigma_\mn P_L \nu \big) \nn\\*
&\qquad\quad + \big(\cbar\, \alTRt \sigma^\mn P_R b\big)
  \big(\bar\ell\, \beTRt \sigma_\mn P_R \nu \big) \big]\,.
\end{align}
\end{subequations}
The NP couplings to the quark and lepton currents are denoted by $\tA^i_j$ and $\tB^i_j$, respectively.
The lower index of $\tB$ denotes the $\nu$ helicity and the lower
index of $\tA$ is that of the $b$ quark.  (This notation is a
variation of the conventions chosen in Ref.~\cite{Ligeti:2016npd}, whence 
the seemingly superfluous tildes. See App.~\ref{app:NPA} for details.) The NP
couplings $\tA^i_j$ and $\tB^k_l$ may be complex, and 
$\tA^i_j \tB^i_l$ products are normalized with respect to
the SM couplings, such that setting $\tA^V_L \tB^V_L = 1$
would amount to doubling the coefficient of the $(\bar c \gamma_\mu P_L
b)(\bar\ell \gamma^\mu P_L \nu)$ operator compared to its SM value. The $CP$
conjugate operators for $\bbar \to \cbar \bar\ell \nu$ are obtained by
Hermitian conjugation.  The operators involving right-handed neutrinos are included for completeness, but do not
interfere with the SM (neglecting $m_\nu$ suppressed terms).  

We define the dimensionless ratios
\beq
r = {m_{\dss}}/{m_B}\,, \qquad \rl = {m_{\ell}^2}/{m_B^2}\,,
\eeq
as well as
\beq
\mSqq = \frac{q^2}{m_B^2} = 1 + \rC^2 - 2 \rC w\,, \qquad
\Gamma_0 = \frac{G_F^2\,|V_{cb}|^2\,m_B^5}{192\pi^3}\,.
\eeq
In the rest of this section, for brevity, we suppress indication of absolute
value squared for the NP$^2$ terms, so that all $(\tA^i_j \tB^i_l)^2$ terms mean
$|\tA^i_j \tB^i_l|^2$.  Similarly, all interference terms are understood as the
appropriate real parts: terms linear in NP couplings (coming from SM--NP
interference) of the form $\tA^i_j \tB^i_k$ mean ${\rm Re}(\tA^i_j \tB^i_l)$,
while bilinear terms in NP couplings (from NP--NP interference) of the form
$\tA^i_j \tB^i_m\, \tA^k_l \tB^k_n$ mean ${\rm Re}(\tA^i_j \tB^i_m\, \tA^k_l{}^*
\tB^k_n{}^*)$.

Considering only left-handed neutrinos, we obtain for the $B\to
D_0^*\ell\bar\nu$ rate, 
\begin{widetext}
\begin{subequations}
\label{eqn:D0srate}
\begin{align}
\frac{\d\Gamma_{D_0^*}^{\rm (SM)}}{\d w}
 & =  4\, \Gamma_0\, r^3 \sqrt{w^2-1}\, \frac{(\mSqq-\rl\big)^2}{\hat q^6}
  \bigg\{ g_-^2 (w-1) \Big[ \rl [(1+r^2) (2 w-1)+2 r (w-2)]
  + (1-r)^2 (w+1) \mSqq \Big] \label{D0rateSM} \\*
&\quad + g_+^2 (w+1) \Big[ \rl [(1+r^2) (2 w+1)-2 r (w+2)]
  +(1+r)^2 (w-1) \mSqq \Big] 
 - 2 g_- g_+ (1-r^2) (w^2-1) (\mSqq+2 \rl) \bigg\}\,. \nn \\
\frac{\d\Gamma_{D_0^*}}{\d w} 
& = \frac{\d\Gamma_{D_0^*}^{\rm (SM)}}{\d w}\,
  \big(1+ \alVLt \beVLt-\alVRt \beVLt\big)^2
  + 2\, \Gamma_0\, r^3 \sqrt{w^2-1}\, \frac{(\mSqq-\rl\big)^2}{\hat q^4}\, \bigg\{ 
  3 \big[(\alSRt - \alSLt) \beSLt\big]^2\, g_P^2\, \mSqq \label{D0rateSM} \\*
&\qquad + 6 (\alSRt - \alSLt) \beSLt g_P \sqrt{\rl}\, 
  \big(1+ \alVLt \beVLt-\alVRt \beVLt\big) 
  \big[g_- (1+r) (w-1) - g_+ (1-r)(w+1)\big] \nn \\*
&\qquad + 8 \alTLt \beTLt g_T (w^2-1) \Big[ 
  2\alTLt \beTLt g_T (\mSqq+2 \rl)
  + 3 \sqrt{\rl}\, \big(1+\alVLt \beVLt-\alVRt \beVLt\big)
  \big[g_+ (1+r) - g_-(1-r)\big] \Big] \bigg\} \,. \nn 
\end{align}
\end{subequations}
For the narrow $D_1$ state ($s_\ell^{\pi_\ell} = \frac32^+$) we find
\begin{subequations}
\label{eqn:D1srate}
\begin{align}
\frac{\d\Gamma_{D_1}^{\rm (SM)}}{\d w} & = 
  2\Gamma_0 r^3 \sqrt{w^2-1}\, \frac{(\mSqq-\rl\big)^2}{\hat q^6} \bigg\{ 
  f_{V_1}^2 \Big[ 2 \mSqq [(w-r)^2+2\mSqq] + \rl \big[4(w-r)^2 - \mSqq\big] \Big] \label{D1rateSM}\\*
&\quad + (w^2-1) \bigg( 
  f_{V_2}^2 \Big[ 2 r^2 \mSqq (w^2-1) + \rl [3\mSqq+4r^2(w^2-1)]\Big]
  + f_{V_3}^2 \Big[2\mSqq (w^2-1) + \rl [4(w-r)^2-\mSqq] \Big] \nn\\*
&\qquad\qquad + 2 f_A^2 \mSqq (2\mSqq+\rl) 
  + 2 f_{V_1} f_{V_2} \Big[ 2 r \mSqq (w-r) + \rl (3-r^2-2 r w) \Big] 
 + 4 f_{V_1}  f_{V_3} (w-r) (\mSqq+2 \rl) \nn\\*
&\qquad\qquad + 2 f_{V_2} f_{V_3} \Big[ 2r\mSqq(w^2-1) + \rl [3w \mSqq + 4r(w^2-1)] \Big]
  \bigg) \bigg\}\,,\nn\\
\frac{\d\Gamma_{D_1}}{\d w} & = 
  \frac{\d\Gamma_{D_1}^{\rm (SM)}}{\d w}\,
  \big(1+ \alVLt \beVLt-\alVRt \beVLt\big)^2
  + 2\Gamma_0 r^3 \sqrt{w^2-1}\, \frac{(\mSqq-\rl\big)^2}{\hat q^6}\, 
  \bigg\{ 3 \big[(\alSLt+\alSRt) \beSLt\big]^2 f_S^2 (w^2-1) \hat q^4 \label{D1rateNP}\\*
&\quad + 6 (\alSLt + \alSRt) \beSLt
  \big(1 + \alVLt \beVLt + \alVRt \beVLt\big)
  f_S (w^2-1) \mSqq \sqrt{\rl}\,
  \big[f_{V_1} + f_{V_2}(1-r w)+f_{V_3} (w-r) \big] \nn\\*
&\quad + 16 (\alTLt \beTLt)^2 (\mSqq+2 \rl) 
  \bigg(f_{T_1}^2 \big[\mSqq(2+w^2) + 4r^2(w^2-1)] 
  + f_{T_2}^2 [4(w-r)^2-\mSqq] + f_{T_3}^2 \mSqq (w^2-1)^2 \nn\\*
&\qquad\qquad + 2 f_{T_1} f_{T_2} [3w \mSqq+4r(w^2-1)]
  - 2 f_{T_3} \big(f_{T_1} w + f_{T_2}\big) \mSqq (w^2-1) \bigg) \nn\\*
&\quad - 24 \alTLt \beTLt \sqrt{\rl} \mSqq \bigg(
  \big(1 + \alVLt \beVLt - \alVRt \beVLt\big)\,
  2 f_A (f_{T_1} r+f_{T_2}) (w^2-1) \nn \\*
&\qquad\qquad - \big(1 + \alVLt \beVLt + \alVRt \beVLt\big)
  \Big[ 2 f_{T_1} f_{V_1} (1-r w)
  + \big[w f_{T_1} + 3 f_{T_2} - f_{T_3}(w^2-1)\big] f_{V_1} (w-r) \nn\\*
&\qquad\qquad + \big[w f_{T_1} + f_{T_2} - f_{T_3} (w^2-1)\big] 
  (f_{V_2} r+f_{V_3})\, (w^2-1) \Big]\bigg) \nn\\*
&\quad + 4 \alVRt \beVLt (1+\alVLt \beVLt) \bigg(
  3 f_{V_1}^2 \mSqq (2 \mSqq + \rl)
  + 2 f_{V_1} \big[f_{V_1} + 2 f_{V_3} (w-r)\big] (w^2-1) (\mSqq+2\rl) \nn\\*
&\qquad\qquad + (w^2-1) \bigg( 
  f_{V_2}^2 \big[ 2 r^2 \mSqq (w^2-1) + \rl [3\mSqq+4r^2(w^2-1)] \big] 
  + f_{V_3}^2 \big[ 2\mSqq (w^2-1) + \rl [4(w-r)^2-\mSqq] \big] \nn\\*
&\qquad\qquad + 2 f_{V_1} f_{V_2} \big[ 2r\mSqq(w-r) + \rl (3-r^2-2 r w)\big]
  + 2 f_{V_2} f_{V_3} \big[ 2r\mSqq(w^2-1) + \rl [3w \mSqq + 4r(w^2-1)] \big]
  \bigg) \bigg) \bigg\} \,. \nn
\end{align}
\end{subequations}
The result for the broad $D_1^*$ state ($s_\ell^{\pi_\ell} = \frac12^+$) can
be obtained from Eqs.~(\ref{D1rateSM}) and (\ref{D1rateNP}) via the
replacements $f_S\to -g_S,\, f_A\to g_A,\, f_{V_i}\to g_{V_i},\, f_{T_i}\to
g_{T_i}$, where $i=1,2,3$. For $\d \Gamma(B\to D_2^*\ell\bar\nu)/\d w$ the result is
\begin{subequations}
\label{eqn:D2rate}
\begin{align}
\frac{\d\Gamma_{D_2^*}^{\rm (SM)}}{\d w} & = 
  \frac{2\Gamma_0}3\, r^3 (w^2-1)^{3/2}\, \frac{(\mSqq-\rl\big)^2}{\hat q^6}
  \bigg\{ k_{A_1}^2 \Big[ 2\mSqq \big[2(w-r)^2 + 3\mSqq\big]
  + \rl [8(w-r)^2-3\mSqq] \Big] \label{D2rateSM}\\*
&\quad + 2 (w^2-1) \bigg( k_{A_2}^2 \Big[ 2 r^2 \mSqq (w^2-1) 
  + \rl \big[3\mSqq + 4r^2 (w^2-1)\big] \Big] 
  + k_{A_3}^2 \Big[2 \mSqq (w^2-1) + \rl [4(w-r)^2-\mSqq] \Big] \nn\\*
&\qquad\qquad + 3 k_V^2 \mSqq (\mSqq+\rl/2) 
  + 2 k_{A_1} k_{A_2} \big[ 2 r\mSqq (w-r) + \rl (3-r^2-2 r w) \big] 
  + 4 k_{A_1} k_{A_3} (w-r) (\mSqq+2 \rl) \nn\\*
&\qquad\qquad + 2 k_{A_2} k_{A_3} \Big[ 2r\mSqq(w^2-1) 
  + \rl \big[3w\mSqq+4r(w^2-1)\big] \Big] \bigg) \bigg\}\,,\nn\\
\frac{\d\Gamma_{D_2^*}}{\d w} & = \frac{\d\Gamma_{D_2^*}^{\rm (SM)}}{\d w}\,
  \big(1+ \alVLt \beVLt-\alVRt \beVLt\big)^2
  + \frac{4\Gamma_0}3\, r^3 (w^2-1)^{3/2} \frac{(\mSqq-\rl\big)^2}{\hat q^6}\, 
  \bigg\{   \label{D2rateNP}\\*
&\quad 6 \alVRt \beVLt (1+\alVLt \beVLt) k_V^2 (w^2-1) \mSqq (2\mSqq+\rl) + 3 \big[(\alSRt - \alSLt) \beSLt\big]^2 k_P^2 (w^2-1) \hat q^4\nn\\*
&\quad + 6 (\alSLt-\alSRt) \beSLt k_P (w^2-1) 
  \mSqq \sqrt{\rl} \big(1 +\alVLt \beVLt-\alVRt \beVLt\big)
  \big[k_{A_1} + k_{A_2} (1-r w)+k_{A_3} (w-r)\big] \nn\\*
&\quad + 16 (\alTLt \beTLt)^2  (\mSqq+2\rl) 
  \bigg( k_{T_1}^2 (w+1) \big[\mSqq(4w+1)+6r(w^2-1)\big] 
  + k_{T_2}^2  (w-1) \big[\mSqq(4w-1)+6r(w^2-1)\big] \nn\\*
&\qquad\qquad + k_{T_3} (w^2-1) \mSqq \big[ k_{T_3} (w^2-1) + 2 k_{T_1}(w+1)
  + 2 k_{T_2}(w-1) \big] - 4 k_{T_1}k_{T_2} (w^2-1) (1+rw-2 r^2) \bigg)
\nn\\*
&\quad + 12 \alTLt \beTLt \sqrt{\rl}\, \mSqq \bigg( (w^2-1)
  \bigg( 2 (k_{A_2}\,r+k_{A_3}) \big(1+\alVLt \beVLt-\alVRt \beVLt\big)\,
  \big[k_{T_1} (w+1) + (w-1) (k_{T_2} + k_{T_3} (1+w)) \big] \nn\\*
&\qquad\qquad - 3 k_V (1 + \alVRt \beVLt + \alVLt \beVLt)
  \big[k_{T_1} (1+r) - k_{T_2} (1-r) \big] \bigg) 
  + k_{A_1} \big(1+\alVLt \beVLt-\alVRt \beVLt\big)\nn\\*
&\qquad\qquad
  \times \big[k_{T_1} (w+1) (3+2w-5 r) - k_{T_2} (w-1) (3-2w+5 r)
  + 2 k_{T_3} (w^2-1)(w-r) \big] \bigg) \bigg\} \,.\nn
\end{align}
\end{subequations}
\end{widetext}
In the heavy quark limit, for the SM and the tensor coupling contributions,
using Eqs.~\eqref{BD12lead} and~\eqref{BD32lead}, our results in
Eqs.~\eqref{eqn:D0srate}--\eqref{eqn:D2rate} agree with the results in Eqs.~(B9)
-- (B12) in Ref.~\cite{Biancofiore:2013ki} (see also
Ref.~\cite{Duraisamy:2014sna}), which studied the $R(\dss)$ predictions for a NP
tensor interaction using QCD sum rule predictions for the leading order
Isgur-Wise functions. In Appendix~\ref{app:DDsrates}, for completeness we
include the analogous expressions for the $B \to \dds$ rates.

In the SM, the only form factors that enter these rates without additional
$(w-1)$ suppressions are $g_+$ for $D_0^*$, $g_{V_1}$ for $D_1^*$, and
$f_{V_1}$ for $D_1$.  In the infinite mass limit, these form factors must
vanish at $w=1$ due to heavy quark symmetry.  The model independent result
derived in the SM is that the ${\cal O}(\lqcd/m_{c,b})$ leading terms for
these form factors at $w=1$ are determined by hadron mass splittings and the
leading order Isgur-Wise functions~\cite{Leibovich:1997em},
\begin{align}\label{modindepSM}
\hat g_+(1) &= -\frac32\, (\varepsilon_c + \varepsilon_b)\, 
  (\bar\Lambda^*-\bar\Lambda) + \ldots \,, \nn\\*
\hat g_{V_1}(1) &= (\varepsilon_c-3\,\varepsilon_b)
  (\bar\Lambda^*-\bar\Lambda) + \ldots \,, \nn\\*
\sqrt6\, \hat f_{V_1}(1) &= 
  - 8\, \varepsilon_c (\bar\Lambda'-\bar\Lambda) + \ldots \,,
\end{align}
where the ellipses denote ${\cal O}[\aS(w-1),\, \varepsilon_{c,b}
(w-1),\, \varepsilon_{c,b}^2]$ and higher order terms.  All terms in
$\d\Gamma_{D_2^*} / \d w$ have an overall $(w^2-1)^{3/2}$ suppression, so there
is no similar constraint for that channel.

In the presence of new physics, the $g_P$ (for $D_0^*$), $g_{T_{1,2}}$ (for
$D_1^*$), and $f_{T_{1,2}}$ (for $D_1$) form factors also have unsuppressed
contributions at $w=1$.  Of these, $g_P(1)$ vanishes in the heavy quark
limit, and is determined at order $\lqcd/m_{c,b}$ by hadron mass splittings
and the leading order Isgur-Wise function similar to Eq.~\eqref{modindepSM}
above
\beq\label{modindepNP1}
\hat g_P(1) = 3\, (\varepsilon_c - \varepsilon_b)
  (\bar\Lambda^*-\bar\Lambda) + \ldots\,.
\eeq
The $f_{T_{1,2}}$ and $g_{T_{1,2}}$ form factors are proportional to the
respective Isgur-Wise functions, $\zeta$ and $\tau$, in the heavy quark
limit, without any $(w-1)$ factors, and their contributions to $\d\Gamma / \d
w$ neither include a $(w-1)$ suppression.  However, in the heavy quark limit,
$g_{T_1} = - g_{T_2}$ and $f_{T_1} = - f_{T_2}$, so only $f_{T_1}+ f_{T_2}$ and 
$g_{T_1}+ g_{T_2}$ terms appear in the rates. One sees in
Eq.~\eqref{D1rateNP} that the sum of the terms proportional to
$f_{T_1}^2$, $f_{T_2}^2$, and $f_{T_1} f_{T_2}$ vanishes at $w=1$ in the
heavy quark limit, as it must (and similarly for the $g_{T_1}^2$,
$g_{T_2}^2$, and $g_{T_1} g_{T_2}$ terms).  The contributions to the rate
proportional to the linear combinations $(f_{T_1} - f_{T_2})$ and $(g_{T_1} -
g_{T_2})$ are suppressed by $(w-1)$,~while
\begin{align}\label{modindepNP2}
\sqrt6\, \big[\hat f_{T_1}(1) + \hat f_{T_2}(1)\big] &=
  -8 \varepsilon_c (\bar\Lambda' - \bar\Lambda) + \ldots\,, \nn\\*
\hat g_{T_1}(1) + \hat g_{T_2}(1) &= (\varepsilon_c + 3\varepsilon_b)
  (\bar\Lambda^* - \bar\Lambda) + \ldots\,,
\end{align}
are again determined at order $\lqcd/m_{c,b}$
by hadron mass splittings and the leading order Isgur-Wise function, similar
to Eqs.~(\ref{modindepSM}) and (\ref{modindepNP1}).

\section{Some predictions}
\label{sec:app}

The data which can be used to constrain the leading and subleading Isgur-Wise
functions are the four $\Bbar\to \dss\l\bar\nu$ branching ratios, the $\Bbar\to
D_2^*\l\bar\nu$ and $\Bbar\to D_0^*\l\bar\nu$ spectra
measured~\cite{Liventsev:2007rb} in four and five bins of $q^2$, respectively,
and the $B\to \dsN\pi$ rates which are related to the semileptonic rates at
$q^2=m_\pi^2$ using factorization.  The available data, used in our fits, is
identical to that collected in Tables~V--VII in Ref~\cite{Bernlochner:2016bci},
and are not repeated here.  In all fits we perform, we expand the leading-order
Isgur-Wise functions in $(w-1)$ to linear order,
\begin{align}
	\label{eqn:LEIW}
	\tau(w) & \simeq \tau(1)\big[1 + \tau'(w-1)\big]\,, \nn\\*
	\zeta(w) & \simeq \zeta(1)\big[1 + \zeta'(w-1)\big]\,.
\end{align}
In our predictions for the $\Bbar \to \dss \tau \nu$ rates and $R(\dss)$, for simplicity we assume that the $l = e, \mu$ rates are given by the SM, and
that NP may only enter the $\tau$ mode. 

\subsection{Fits in the heavy quark limit}

\begin{table}[b]
\renewcommand*{\arraystretch}{1.2}
\newcolumntype{C}{ >{\centering\arraybackslash $} c <{$}}
\begin{tabular}{C|CC}
	       \hline\hline
	       \chi^2/\text{dof} &\tau(1) & \tau' \\\hline
       2.5/3 & 0.65 \pm 0.08 & -1.3 \pm  0.4   \\\hline\hline
\tau(1)        & 1 & -0.90 \\
\tau'	       & -0.90 & 1 \\
\hline
\end{tabular}

\vspace{10pt}
\begin{tabular}{C|CC}
	       \hline\hline
	       \chi^2/\text{dof} &\zeta(1) & \zeta' \\\hline
       9.1/4   &1.14 \pm   0.32 & -0.20  \pm   1.0\\\hline\hline
\zeta(1)       & 1 & -0.95 \\
\zeta'         & -0.95 & 1 \\
\hline
\end{tabular}
\caption{Fit results and correlations, fitting at leading order in HQET the
$\Bbar\to D_2^* l\bar\nu$ data (above) and $D_0^*l\bar\nu$ (below).}
\label{tab:LOfit}
\end{table}

To understand the importance of the $\mathcal{O}(\lqcd/m_{c,b})$ corrections, 
it is instructive to attempt fitting the data using the form factor
parametrizations in the heavy quark limit. Fitting the $\d \Gamma/\d w$ data for
$\dTs$ and $\dSs$ and the branching fractions of all four states in the heavy
quark limit, using Eqs.~\eqref{BD12lead} and \eqref{BD32lead}, yields an
unacceptably poor fit ($\chi^2/\text{dof} = 80./4$).  The fit is not improved
with the addition of quadratic terms to~Eqs.~\eqref{eqn:LEIW}.  Given the
branching ratios and some gain in efficiencies in Belle~II over Belle, one may
expect sensitivity to such quadratic terms with about (5--10)/ab of Belle~II
data.

To better quantify this tension, we instead fit $\tau(1)$ and $\tau'$
($\zeta(1)$ and $\zeta'$) using the branching ratio and $\d\Gamma/\d w$ for the
$\dTs$ ($\dSs$) alone, and the $\Bbar \to \dTs\pi$ rate. This yields good fits,
with the results shown in Table~\ref{tab:LOfit}.  From these fits one can
predict the ratios,
\begin{align}
R_{3/2} & = \frac{\Gamma[\Bbar \to \dTs l \bar\nu]}{\Gamma[\Bbar \to \dV l \bar\nu]} \simeq 1.67 \pm 0.09\,,\nn \\
R_{1/2} & = \frac{\Gamma[\Bbar \to \dSs l \bar\nu]}{\Gamma[\Bbar \to \dVs l \bar\nu]} \simeq  0.88\pm 0.07 \,,
\end{align}
to be compared to the current experimental values~\cite{Liventsev:2007rb,
Bernlochner:2016bci}, $R_{3/2} \simeq 0.45 \pm 0.07$ and $R_{1/2} \simeq 2.2 \pm
0.7$.  (The $R_{3/2} = 1.67$ central value is very close to the `$B_\infty$'
result, 1.65, in Table~II in Ref.~\cite{Leibovich:1997em}.)  The severe tension for
$R_{3/2}$, about $10\sigma$, is evidence for the presence of large deviations
from the heavy quark limit.  Adding quadratic terms to Eqs.~\eqref{eqn:LEIW}
does not resolve this tension. However, including the  $\mathcal{O}(\lqcd/m_{c,b})$ corrections yields 
good fits~\cite{Bernlochner:2016bci}, in alignment with the expectation that these corrections 
can be large because of the zero recoil suppression of the heavy quark limit terms. Therefore, hereafter we
consider only fits including both $\lqcd/m_{c,b}$ and $\aS$ terms.

\subsection{Fits including the subleading terms}

To deal with the several unknown subleading Isgur-Wise functions and the limited
amount of experimental data, some approximations need to be made.  In what is
called ``Approximation~A'' in Ref.~\cite{Leibovich:1997em}, $(w-1)$ is treated
as a small parameter of order $\lqcd/m_{c,b}$, up to second order terms are
kept, and chromomagnetic terms are neglected.  This reduces the number of
subleading Isgur-Wise functions that enter, and allows parametrization of the
rates with a few numbers (rather than functions).  In ``Approximation~B'' one
does not expand in $(w-1)$, but still neglects chromomagnetic terms.  Finally,
Ref.~\cite{Bernlochner:2016bci} introduced an ``Approximation~C'', which treats
$\hat\tau_{1,2}$ and $\hat\zeta_1$ as constant fit parameters. It also does not
neglect the subleading Isgur-Wise functions parametrizing matrix elements of
the chromomagnetic term in the Lagrangian, motivated by the fact that the
$m_{\dVs} - m_{\dSs}$ mass splitting no longer appears much smaller than
$m_{D^*} - m_D$. 

We have updated the Approximation~C fit to include the $\alpha_s$ corrections
neglected in Ref.~\cite{Bernlochner:2016bci}. This changes the fit parameters
shown in Table~\ref{tab:AppB_results} only slightly compared to
Ref.~\cite{Bernlochner:2016bci}.  For $R(\dss)$ defined in Eq.~(\ref{RXdef}) we
obtain
\begin{align}\label{eq:AppB_RDds_results}
	R(\dSs) &= 0.08 \pm 0.03\,,  &  \widetilde R(\dSs) &= 0.24 \pm 0.05\,,\nn\\*
	R(\dVs) &= 0.05 \pm 0.02\,,  &  \widetilde R(\dVs) &= 0.18 \pm 0.02\,, \nn \\*
 	R(\dV) &=  0.10 \pm 0.02\,,  &  \widetilde R(\dV) &= 0.20 \pm 0.02\,, \nn \\*
 	R(\dTs) &= 0.07 \pm 0.01\,,  &   \widetilde R(\dTs) &= 0.17\pm 0.01\,,
\end{align}
and the phase-space constrained ratio is defined as
\beq
\widetilde R(X) = \frac{\ds \int_{m_\tau^2}^{(m_B-m_X)^2} 
  \frac{\d\Gamma(\Bbar \to X \tau\bar\nu)}{\d q^2}\, \d q^2 }
  {\ds \int_{m_\tau^2}^{(m_B-m_X)^2} 
  \frac{\d\Gamma(\Bbar \to X l\bar\nu)}{\d q^2}\, \d q^2} \,.
\eeq
Our results in Eq.~(\ref{eq:AppB_RDds_results}) are nearly identical with
Eq.~(38) in Ref.~\cite{Bernlochner:2016bci}. 
The ratio for the sum of the four $\dss$ states~is
\begin{equation}
\ov{\!R}(\dss) = \frac{\sum_{X \in \dss} \Gamma[\Bbar \to X \tau \bar\nu]}
  {\sum_{X \in \dss} \Gamma[\Bbar \to X l \bar\nu]} = 0.08 \pm 0.01\,.
\end{equation}

\begin{table}[t]
\renewcommand*{\arraystretch}{1.2}
\newcolumntype{C}{ >{\centering\arraybackslash $} c <{$}}
\begin{tabular}{C|CCCC}
	\hline\hline
	\chi^2/\text{dof} &\tau(1) & \tau' & \hat\tau_1 & \hat\tau_2  \\\hline
	2.4/4	&0.70 \pm 0.07 & -1.6 \pm 0.2  & -0.5 \pm 0.3  & 2.9 \pm 1.4 \\ \hline\hline
\tau(1) 	& 1 & -0.85 & 0.53 & -0.49 \\
\tau' 		& -0.85 & 1 & -0.17 & 0.086 \\
\hat\tau_1	&  0.53 & -0.17 & 1 & -0.89 \\
\hat\tau_2	& -0.49 & 0.086 & -0.89 & 1 \\
\hline
\end{tabular}

\vspace{10pt}
\begin{tabular}{C|CCC}
	\hline\hline
		\chi^2/\text{dof} & \zeta(1) & \zeta' & \hat\zeta_1  \\ \hline
	9.1/4	& 0.70 \pm 0.21 & 0.2 \pm 1.4 & 0.6 \pm 0.3 \\ \hline\hline
\zeta(1)  	& 1 & -0.95 & -0.44 \\
\zeta'  	& -0.95 & 1 & 0.61 \\
\hat\zeta_1	& -0.44 & 0.61 & 1 \\
\hline
\end{tabular}
\caption{Fit results and correlations for Approximation~C, for the narrow
$\dsN$ (above) and broad $\dsW$ (below) states.}
\label{tab:AppB_results}
\end{table}

\begin{figure*}[t]
\includegraphics[width=\columnwidth]{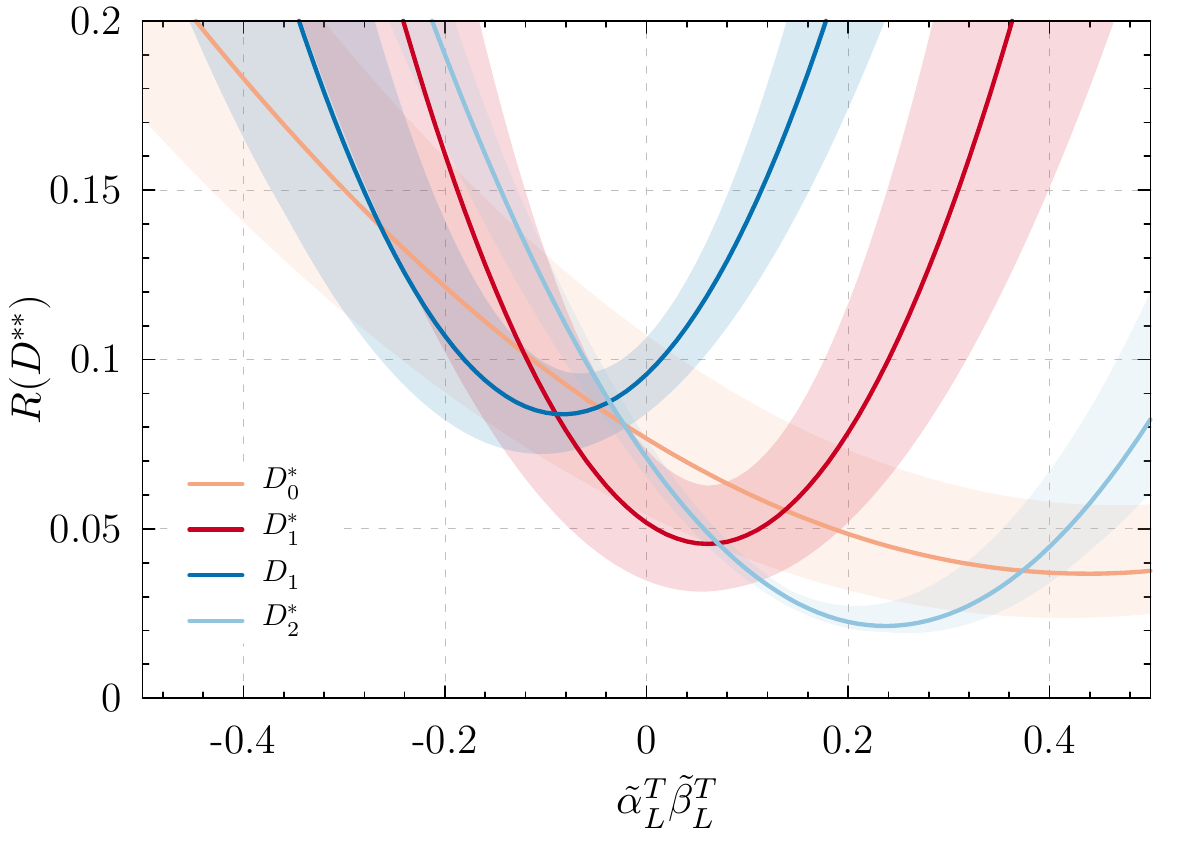} \hfill
\includegraphics[width=\columnwidth]{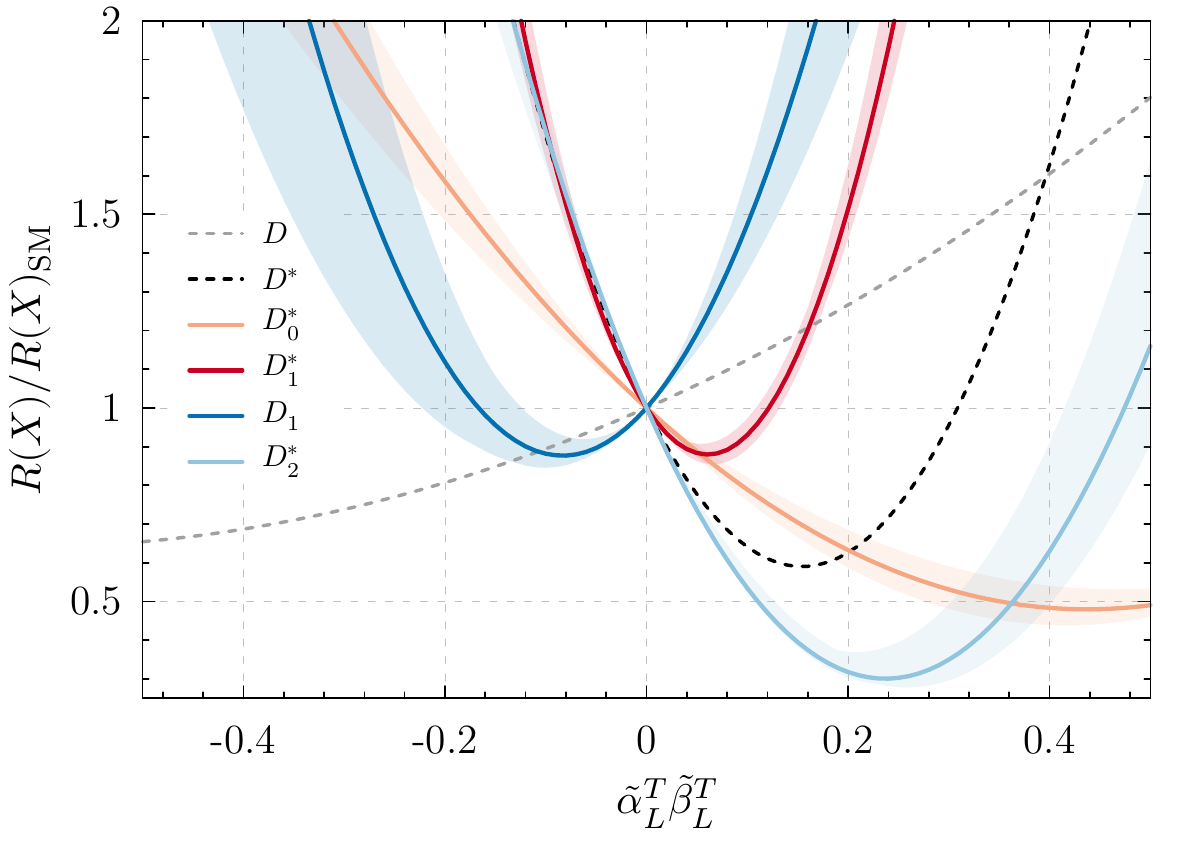}
\caption{Predictions for $R(D^{**})$ as functions of a NP tensor coupling
(left), and for $R(X) / R(X)_{\rm SM}$ for all six $D^{(*,**)}$ states~(right).}
\label{fig:tensor}
\end{figure*}

We next consider new physics generating one of the interactions in the basis
defined by Eq.~\eqref{eqn:Odef}.  The scalar ($O_S$) and pseudoscalar ($O_P$)
matrix elements could be estimated in the past using
Eq.~\eqref{currentrel}~\cite{Bernlochner:2016bci}.  A right-handed vector
current, $O_V+O_A$, does not help to fit the current data, and $O_V-O_A$ is the
SM operator.  Hence, we study here in some detail NP generating the tensor
operator, $O_T$, which has best fit for the present data at $\alTLt\beTLt \simeq
0.35$, assuming it is real (corresponding to $C_T \simeq 0.48$ in the
conventions of Ref.~\cite{Freytsis:2015qca}).  This is for primarily for
illustration: While exclusively $O_T$ cannot be generated by a dimension-6 new
physics operator, it can arise from Fierzing interactions generated in viable
scenarios.

\begin{figure}[t]
\includegraphics[width=\columnwidth]{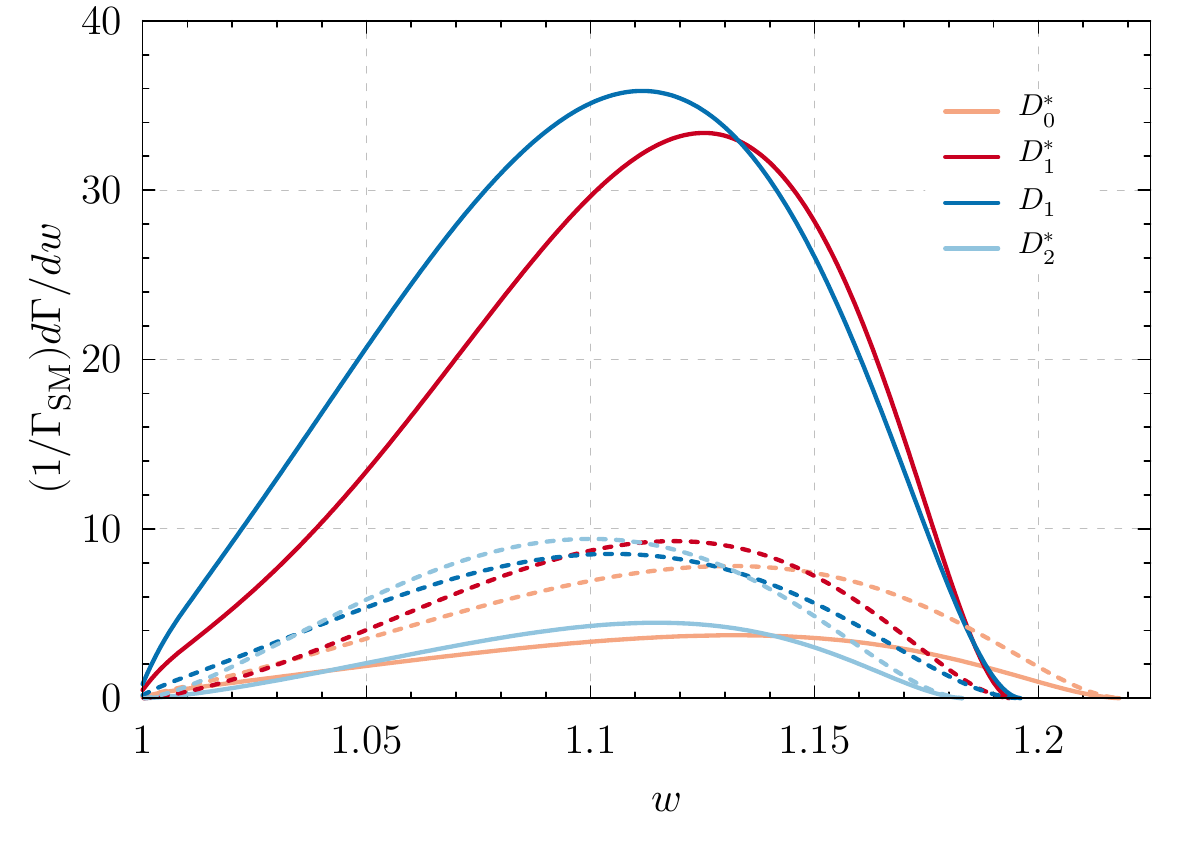}
\caption{Predicted $\d \Gamma/\d w$ distributions normalized to the SM rates for
each of the four $D^{(**)}$ states, in the SM (dashed curves) and for
$\alTLt\beTLt = 0.35$ (solid curves), as determined from the Approximation~C best
fit result, including $\lqcd/m_{c,b}$ and $\alpha_s$ corrections.}
\label{fig:wSpectra}
\end{figure}

Figure~\ref{fig:tensor} (left) shows our predictions for
$R(D^{**})$ as functions of $\alTLt\beTLt$, and Fig.~\ref{fig:tensor} (right) shows the predictions for $R(X) / R(X)_{\rm
SM}$ for all six $D^{(*,**)}$ states.  In the vicinity of $\alTLt\beTLt \simeq
0.35$, where the current $R(D)$ and $R(D^*)$ data can be fit well, measurements
of $R(D^{**})$ will have a lot of discriminating power.  We obtain for
$\alTLt\beTLt = 0.35$ the central values for $R(X) / R(X)_{\rm SM} = \{0.46,\,
4.3,\, 4.3,\, 0.47\}$ for $\{ D_0^*,\, D_1^*,\, D_1,\, D_2^*\}$, respectively,
whereas the corresponding values for $\{ D,\, D^*\}$ are $\{1.51,\, 1.25\}$. 
The uncertainty bands are dominated by the first principal components of the fit
covariance matrices, added in quadrature with variations in $\zeta_1$ and
$\tau_1$~(see Ref.~\cite{Bernlochner:2016bci} for details).
Figure~\ref{fig:wSpectra} shows the predicted $\d \Gamma/\d w$ spectra both for
the SM and for $\alTLt\beTLt \simeq 0.35$.

Understanding $\Bbar\to \dss l\bar\nu$ is also important because they give some
of the largest experimental backgrounds to the measurements of $R(D^{(*)})$. 
For example, for Belle, Table~IV in Ref.~\cite{Huschle:2015rga} showed that the
$\Bbar\to D^{**}\ell\bar\nu$ shapes and composition are significant backgrounds
to the $R(D^{(*)})$ measurements.  For BaBar, Table~V in
Ref.~\cite{Lees:2013uzd} lists separately the uncertainties due to $\Bbar\to
D^{**} l\bar\nu$ and $\Bbar\to D^{**} \tau\bar\nu$, which are both significant. 
The sensitive dependence of the $\Bbar\to D^{**} \ell \bar\nu$ rates on
$\alTLt\beTLt$ shown in Figs.~\ref{fig:tensor} and~\ref{fig:wSpectra} illustrate
the importance of treating these $\dss$ backgrounds to $\Bbar\to D^{(*)} \ell
\bar\nu$ consistently, when fitting NP to more precise future
$R(D^{(*)})$~data. 

\begin{figure}[t]
\includegraphics[width=\columnwidth]{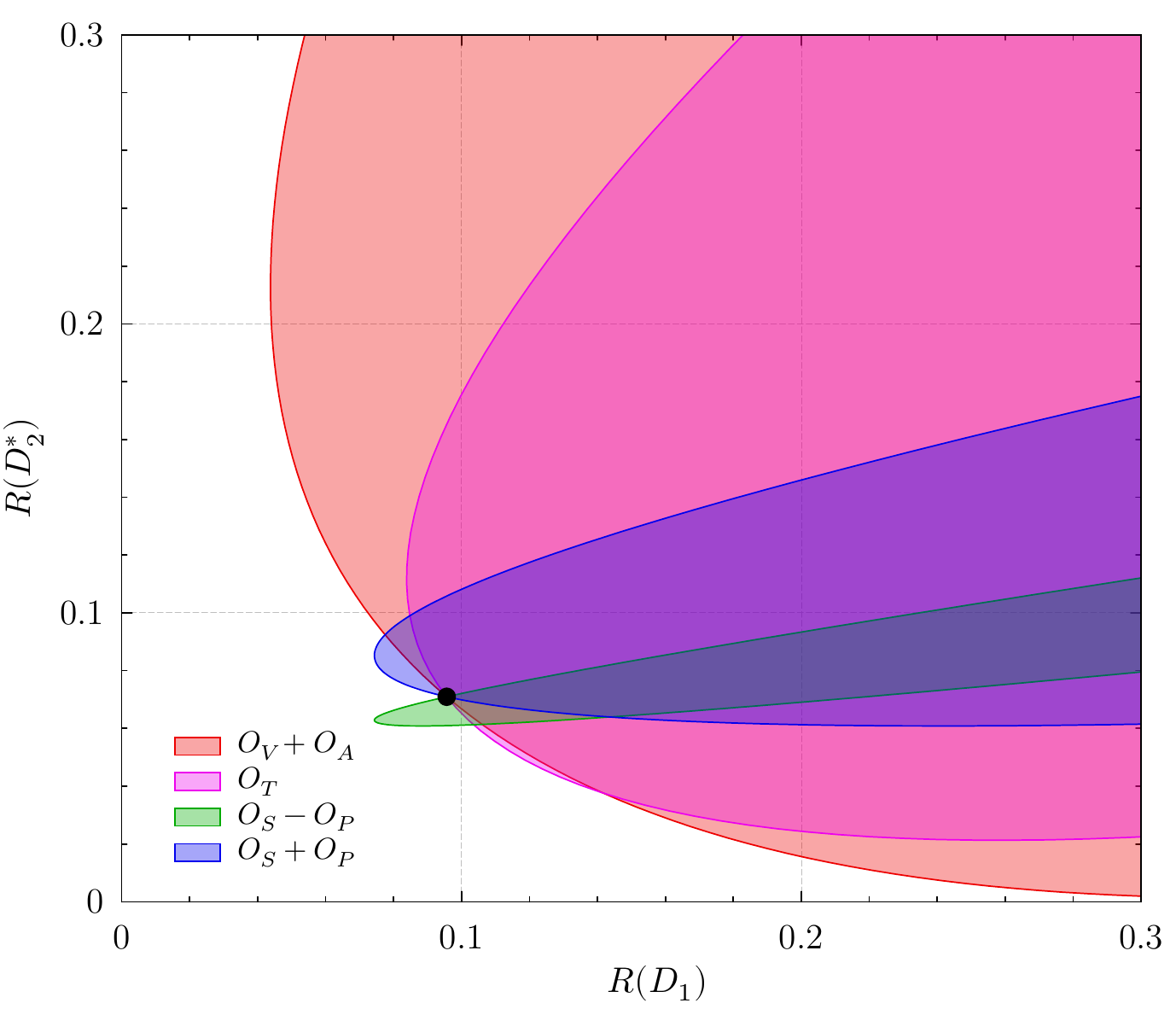}
\caption{Allowed ranges of $R(D_1)$ and $R(D_2^*)$ for the two narrow states, in
the presence of any one of the four non-SM currents with arbitrary weak phases, as determined from the Approximation~C best
fit result, including $\lqcd/m_{c,b}$ and $\alpha_s$ corrections. The SM prediction is shown by the black dot.}
\label{fig:narrowNP}
\end{figure}

Figure~\ref{fig:narrowNP} shows the possible ranges of $R(D_1)$ and $R(D_2^*)$,
for the two narrow states, allowing any one of the four non-SM interactions,
with arbitrary relative phases compared to the SM.  The Approximation~C best
fit result is used, including $\lqcd/m_{c,b}$ and $\alpha_s$ corrections.

Predictions for $\Bbar_s \to D_s^{**} \ell\bar\nu$ can be made using the same
formalism with the appropriate hadron masses, the $\bar\Lambda^{(\prime,*)}$
parameters in Table~\ref{tab:input_summary}, and flavor $SU(3)$ symmetry for the
form factors.  Three of the $D_s^{**}$ states have widths below a few MeV, and
may become the first $\Bbar_{(s)} \to \dss_{(s)}\tau\bar\nu$ decay modes
measured by LHCb.  The $R(D_s^{**})$ predictions are numerically close to those
for $\Bbar \to \dss \ell\bar\nu$, but larger uncertainties arise from $SU(3)$
violation and questions remain regarding the interpretation of $D_s^{**}$ as
simply orbitally excited $\bar s c$ states~\cite{Bernlochner:2016bci}.

\section{Conclusions}
\label{sec:concl}

We derived the $\Bbar\to \dss\ell\bar\nu$ decay rates for arbitrary
beyond SM $b\to c$ currents and finite charged lepton mass, including all order
$\lqcd/m_{c,b}$ and $\alpha_s$ terms in the heavy quark effective theory
expansion of the form factors.  

To describe all $b\to c$ current matrix elements, including $\lqcd/m_{c,b}$ and
$\alpha_s$ corrections, only four functions of $w$  ($\zeta$, $\hat\zeta_1$, and
$\hat\chi_{1,2}$) are needed to determine all twelve $B\to \dsW$ form factors in
Eqs.~\eqref{BD0m} and~\eqref{BD1sm} to this order, as well as the mass
parameters $\bar\Lambda^*$ and $\bar\Lambda$.  For $B\to \dsN$ decays, six
functions of $w$ ($\tau$, $\hat\tau_{1,2}$, and $\hat\eta_{1,2,3}$) describe
the sixteen form factors in Eqs.~(\ref{BD1m}) and (\ref{BD2m}) at this order,
plus the mass parameters $\bar\Lambda'$ and $\bar\Lambda$.  

With the above results, we have now all ingredients in place to consistently
study semileptonic $B$ decays to the six lightest charm mesons, $\Bbar\to
D^{(*,**)} \ell\bar\nu$, for arbitrary new physics and for arbitrary charged
lepton masses.  These results are being implemented in the {\tt
Hammer}~\cite{Duell:2016maj} analysis software, which will allow reweighing
fully simulated data for fully differential decays to arbitrary new physics, including arbitrary NP 
contributions for each of the three lepton flavors.  This
will lead to better control of both theoretical and experimental uncertainties
in $R(D^{(*,**)})$ measurements, as well as in the determinations of $|V_{cb}|$
and $|V_{ub}|$ from semileptonic $B$ decays.

Unlike calculations using model-dependent inputs on the form factors, our
predictions are systematically improvable with more data on the $\Bbar\to \dss
l\bar\nu$ decays to light lepton final states and/or input from lattice QCD. 
The upcoming much larger data sets at LHCb and Belle~II will answer many
important questions.

\acknowledgments

We thank Stephan Duell and Michele Papucci for helpful conversations.
We thank the Aspen Center of Physics, supported by the NSF grant 
PHY-1066293, where parts of this work were completed. 
FB was supported by the DFG Emmy-Noether Grant No.\ BE~6075/1-1. 
ZL was supported in part by the U.S.\ Department of Energy under
contract DE-AC02-05CH11231. 
DR acknowledges support from the University of Cincinnati.

\appendix
\section{NP Amplitudes}
\label{app:NPA}
In this appendix we provide explicit results for the $B \to \Dbar^{**} \bar\ell \nu$ 
amplitudes themselves. These $B \to \Dbar^{**}$ amplitudes correspond to those used in the \texttt{hammer} code~\cite{Duell:2016maj}.

As in Ref.~\cite{Ligeti:2016npd}, we consider the $\bbar \to \cbar$ amplitudes, defining the basis of NP operators to be
\begin{subequations}\label{abdef}
\begin{align}
\text{SM:\,} & \phantom{-}i2\sqrt{2}\, V_{cb}^* G_F\big[\bbar \g^\mu P_L c\big] \big[\bar\nu \g_\mu P_L \ell\big]\,, \\*
\text{Vector:\,} & \phantom{-}i2\sqrt{2}\, V_{cb}^* G_F
  \big[\bbar\big(\alVL \g^\mu P_L + \alVR \g^\mu P_R\big)c\big] \nn\\*
&\qquad\quad \times
  \big[\bar\nu\big(\beVL \g_\mu P_L + \beVR \g_\mu P_R\big) \ell\big]\,, \\
\text{Scalar:\,} & -i2\sqrt{2}\, V_{cb}^* G_F
  \big[\bbar\big(\alSL P_L + \alSR P_R\big)c\big] \nn\\*
&\qquad\quad \times
  \big[\bar\nu\big(\beSL P_R + \beSR P_L\big) \ell\big]\,, \\
\text{Tensor:\,} & -i2\sqrt{2}\, V_{cb}^* G_F
  \big[ \big(\bbar \alTR \sigma^\mn P_R c\big)
  \big(\bar\nu \beTL \sigma_\mn P_R \ell \big) \nn\\*
&\qquad\quad + \big(\bbar \alTL \sigma^\mn P_L c\big)
  \big(\bar\nu \beTR \sigma_\mn P_L \ell \big) \big]\,.
\end{align}
\end{subequations}
The lower index of $\beta$ denotes the $\nu$ helicity and the lower
index if $\alpha$ is that of the $c$ quark.
The correspondence between these coefficients and those defined in
Eq.~(\ref{abtdef}) is (equivalent to)
\begin{align}
\alVL &= \alVLt{}^*\,, & \alVR & = \alVRt{}^* \,, \nn\\*
\alSL &= -\alSRt{}^*\,, & \alSR &= -\alSLt{}^* \,, \nn\\*
\alTL &= -\alTRt{}^*\,, & \alTR &= -\alTLt{}^* \,, \nn\\*
\beta^i_j &= \tB^i_j{}^*\,,&&
\end{align}
Operators for the CP conjugate $b \to c$ processes are obtained by Hermitian conjugation. 

The $B \to \Dbar^{**} \bar\ell \nu$ process features only a single physical
polar helicity angle, $\thtau$, defined in Fig.~\ref{fig:HAD} below.
(Helicity angles and momenta are defined with respect to the $\bbar \to
\cbar$ process. Definitions for the conjugate process follow by replacing all
particles with their antiparticles.) With respect to the $\dss \to D Y$ decay
products, one may define $\phtau$ and $\phD$ as twist angles of the
$\ell$--$\nu$ and $D$--$Y$ decay planes, in accordance with
Fig.~\ref{fig:HAD}: The combination $\phtau - \phD$ becomes a physical phase in the
presence of $\dss$ decays.
Anticipating the need to account for interference effects once
$\dss$ and $\tau$ decays are included, we therefore write results for the
helicity amplitudes including this physical phase combination.  We adopt
conventions that match the spinor conventions of Ref.~\cite{Ligeti:2016npd}
for the $\tau$ decay amplitudes. For left-handed neutrino amplitudes, this is
achieved by including an extra $\dss$ phase factor $e^{i\lambda_{\dss} \phD}$
for $\dss$ spin $\lambda_D$, and an additional spinor phase function,
$h_{s_\ell} = 1$, $e^{-i\phtau}$ for $s_\ell = 1$, $2$ respectively. (We
label the leptonic spin by $1$ and $2$, to distinguish it from the $\dss$
spins as well as to match the conventions of Ref.~\cite{Ligeti:2016npd} for
massive spinors.) 

\begin{figure*}[tb]
	\includegraphics[width = .75\textwidth]{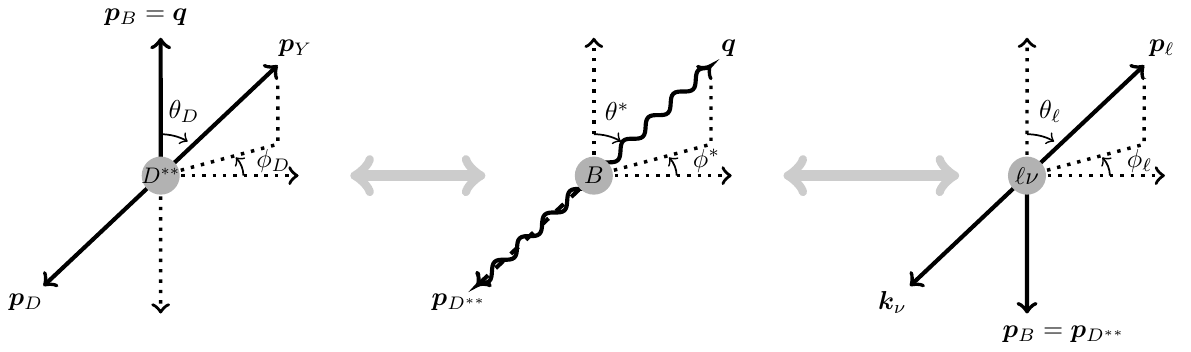}
	\caption{Helicity angle definitions}
	\label{fig:HAD}
\end{figure*}

Under these conventions, the fully differential $B \to \Dbar^{**} \bar\ell
\nu$ rates may be written as
\begin{align}
\frac{\d^2 \Gamma}{\d w\, \d\cos\thtau} &= \frac{G_F^2 m_B^5}{64\pi^3}\,
  \frac{2\rC^3 \sqrt{w^2-1}\, \big(\mSqq-\rt^2\big)^2}{\mSqq} \nn\\*
&\quad \times \sum_{\lambda_{\dss}, s_\ell} \Big| A^{\lambda_{\dss}}_{s_\ell} \Big|^2\,,
\end{align}
in which $r_\ell = m_\ell/m_B = \sqrt{\rho_\ell}$.

In the following subsections, we write the $A_{s_\ell}^{\lambda_{\dss}}$
helicity amplitudes for the $\bbar \to \cbar$ processes, for left-handed
neutrinos only. As in Ref.~\cite{Ligeti:2016npd} , the amplitudes for
right-handed neutrinos are straightforward to include. The $CP$ conjugate $b
\to c$ processes are obtained via
\begin{equation}
	\mathcal{A}_{b \to c}^s(\theta,\phi; \alpha, \beta) = \mathcal{A}_{\bbar \to \cbar}^{\bar{s}}(\theta, -\phi; \alpha^*, \beta^*)\,,
\end{equation}
where $s$ is the set of external state quantum numbers, and $\bar{s}$ denotes
their corresponding $CP$ conjugates. For the $\dTs$ processes, the
$\lambda_{\dTs} = \pm 2$ amplitudes are all zero, and are not included. We
label the remaining amplitudes via $\lambda_{\dTs} = \pm$, $0$.

\vspace*{1cm}

\begin{widetext}
\subsubsection{\texorpdfstring{$B \to \Dbar_0^* \ell \nu$}{D0s}}

\begin{align}
A_1/\sqrt{2} & =
 \bigg\{ \frac{1}{2} g_P  (\alSL -\alSR ) \beSL  \rS+2 g_T  \sqrt{w^2-1} \alTR  \beTL  \rT \cos\thtau  \notag \\
    & -\frac{g_+  \rt (1 + (\alVL - \alVR)\beVL \rV) \big((\rC-1) (1+w)+(1+\rC) \sqrt{w^2-1} \cos\thtau \big)}{2 \mSqq } \notag \\
    & -\frac{g_-  \rt (1 + (\alVL - \alVR)\beVL \rV) \big((1+\rC) (w-1)+(\rC-1) \sqrt{w^2-1} \cos\thtau \big)}{2 \mSqq }\bigg\}\notag \\ 
A_2/\sqrt{2} & =
 \sin\thtau \bigg\{ -2 g_T  \rt \sqrt{\frac{w^2-1}{\mSqq }} \alTR  \beTL  \rT  +\frac{1}{2} g_+  (1+\rC) \sqrt{\frac{w^2-1}{\mSqq }} (1 + (\alVL - \alVR)\beVL \rV) \notag \\ 
    & +\frac{1}{2} g_-  (\rC-1) \sqrt{\frac{w^2-1}{\mSqq }} (1 + (\alVL - \alVR)\beVL \rV)\bigg\}\notag\,.
\end{align}

\subsubsection{\texorpdfstring{$B \to  \Dbar_1 \ell \nu$}{D1}}

\begin{align}
A_1^{-} & =
 \sin\thtau  e^{-i\phDtau}\bigg\{ \frac{1}{2} f_A  \rt \sqrt{\frac{w^2-1}{\mSqq }} (1 + (\alVL - \alVR)\beVL \rV)  +\frac{f_{V_1}  \rt (1 + (\alVL + \alVR)\beVL \rV)}{2 \sqrt{\mSqq }} \notag \\ 
    & +\frac{2 \big(f_{T_2}  \big(\rC-w+\sqrt{w^2-1}\big)+f_{T_1}  (-1 + \rC w + \rC \sqrt{w^2-1}) \alTR  \beTL  \rT}{\sqrt{\mSqq }}\bigg\}\notag \\ 
A_2^{-} & =
 \cos^2\frac{\thtau}{2} e^{-i\phDtau}\bigg\{ f_A  \sqrt{w^2-1} (1 + (\alVL - \alVR)\beVL \rV)  +f_{V_1}  (1 + (\alVL + \alVR)\beVL \rV) \notag \\ 
    & +\frac{4 \rt \big(f_{T_2}  \big(\rC-w+\sqrt{w^2-1}\big)+f_{T_1}  (-1 + \rC w + \rC \sqrt{w^2-1}) \alTR  \beTL  \rT}{\mSqq }\bigg\}\notag \\ 
A_1^{0} & =
 \bigg\{ -\frac{f_S  \sqrt{w^2-1} (\alSL +\alSR ) \beSL  \rS}{\sqrt{2}}  +\frac{f_{V_1}  \rt (1 + (\alVL + \alVR)\beVL \rV) \big(\sqrt{w^2-1}+(\rC-w) \cos\thtau \big)}{\sqrt{2} \mSqq } \notag \\ 
    & -\frac{\rt \sqrt{w^2-1} (1 + (\alVL + \alVR)\beVL \rV) \big(f_{V_3}  (\rC-w)+f_{V_2}  (\rC w -1)+(f_{V_3} +f_{V_2}  \rC) \sqrt{w^2-1} \cos\thtau \big)}{\sqrt{2} \mSqq } \notag \\ 
    & +2 \sqrt{2} \big(f_{T_2} +f_{T_3} +f_{T_1}  w-f_{T_3}  w^2\big) \alTR  \beTL  \rT \cos\thtau \bigg\}\notag \\ 
A_2^{0} & =
 \sin\thtau \bigg\{ -\frac{f_{V_1}  (\rC-w) (1 + (\alVL + \alVR)\beVL \rV)}{\sqrt{2} \sqrt{\mSqq }}  +\frac{(f_{V_3} +f_{V_2}  \rC) \big(w^2-1\big) (1 + (\alVL + \alVR)\beVL \rV)}{\sqrt{2} \sqrt{\mSqq }} \notag \\ 
    & -\frac{2 \sqrt{2} \rt \big(f_{T_2} +f_{T_3} +f_{T_1}  w-f_{T_3}  w^2\big) \alTR  \beTL  \rT}{\sqrt{\mSqq }}\bigg\}\notag \\ 
A_1^{+} & =
 \sin\thtau  e^{+i\phDtau}\bigg\{ \frac{1}{2} f_A  \rt \sqrt{\frac{w^2-1}{\mSqq }} (-1 + (\alVR - \alVL)\beVL \rV)  +\frac{f_{V_1}  \rt (1 + (\alVL + \alVR)\beVL \rV)}{2 \sqrt{\mSqq }} \notag \\ 
    & -\frac{2 \big(f_{T_2}  (w -\rC + \sqrt{w^2-1})+f_{T_1}  (1 - \rC w + \rC \sqrt{w^2-1}) \alTR  \beTL  \rT}{\sqrt{\mSqq }}\bigg\}\notag \\ 
A_2^{+} & =
 \sin^2\frac{\thtau}{2} e^{+i\phDtau}\bigg\{ f_A  \sqrt{w^2-1} (1 + (\alVL - \alVR)\beVL \rV)  +f_{V_1}  (-1 - (\alVL + \alVR)\beVL \rV) \notag \\ 
    & +\frac{4 \rt \big(f_{T_2}  (w -\rC + \sqrt{w^2-1})+f_{T_1}  (1 - \rC w + \rC \sqrt{w^2-1}) \alTR  \beTL  \rT}{\mSqq }\bigg\}\notag\,.
\end{align}

\subsubsection{\texorpdfstring{$B \to \Dbar_2^* \ell \nu$}{D2s}}

\begin{align}
A_1^{-} & =
 \sin\thtau  e^{-i\phDtau}\bigg\{ \frac{k_V  \rt \big(w^2-1\big) (1 + (\alVL + \alVR)\beVL \rV)}{2 \sqrt{2} \sqrt{\mSqq }}  +\frac{k_{A_1}  \rt \sqrt{w^2-1} (1 + (\alVL - \alVR)\beVL \rV)}{2 \sqrt{2} \sqrt{\mSqq }} \notag \\ 
    & +\frac{\sqrt{2} k_{T_1}  (1+w) \big((1+\rC) (w-1)-(1-\rC) \sqrt{w^2-1}\big) \alTR  \beTL  \rT}{\sqrt{\mSqq }} \notag \\ 
    & +\frac{\sqrt{2} k_{T_2}  (w-1) \big((\rC-1) (1+w)+(1+\rC) \sqrt{w^2-1}\big) \alTR  \beTL  \rT}{\sqrt{\mSqq }}\bigg\}\notag \\ 
A_2^{-} & =
 \cos^2\frac{\thtau}{2} e^{-i\phDtau}\bigg\{ \frac{k_V  \big(w^2-1\big) (1 + (\alVL + \alVR)\beVL \rV)}{\sqrt{2}} +\frac{k_{A_1}  \sqrt{w^2-1} (1 + (\alVL - \alVR)\beVL \rV)}{\sqrt{2}} \notag \\ 
    & +\frac{2 \sqrt{2} k_{T_1}  \rt (1+w) \big((1+\rC) (w-1)-(1-\rC) \sqrt{w^2-1}\big) \alTR  \beTL  \rT}{\mSqq } \notag \\ 
    & +\frac{2 \sqrt{2} k_{T_2}  \rt (w-1) \big((\rC-1) (1+w)+(1+\rC) \sqrt{w^2-1}\big) \alTR  \beTL  \rT}{\mSqq }\bigg\}\notag \\ 
A_1^{0} & =
 \bigg\{ \frac{k_P  \big(w^2-1\big) (\alSL -\alSR ) \beSL  \rS}{\sqrt{3}} \notag \\ 
    & +\frac{k_{A_1}  \rt (1 + (\alVL - \alVR)\beVL \rV) \big(w^2-1+(\rC-w) \sqrt{w^2-1} \cos\thtau \big)}{\sqrt{3} \mSqq } \notag \\ 
    & -\frac{\rt \big(w^2-1\big) (1 + (\alVL - \alVR)\beVL \rV) \big(k_{A_3}  (\rC-w)+k_{A_2}  (\rC w -1)+(k_{A_3} +k_{A_2}  \rC) \sqrt{w^2-1} \cos\thtau \big)}{\sqrt{3} \mSqq } \notag \\ 
    & +\frac{4 k_{T_1}  (1+w) \sqrt{w^2-1} \alTR  \beTL  \rT \cos\thtau }{\sqrt{3}} +\frac{4 k_{T_2}  (w-1) \sqrt{w^2-1} \alTR  \beTL  \rT \cos\thtau }{\sqrt{3}} \notag \\ 
    & +\frac{4 k_{T_3}  \big(w^2-1\big)^{3/2} \alTR  \beTL  \rT \cos\thtau }{\sqrt{3}}\bigg\}\notag \\ 
A_2^{0} & =
 \sin\thtau \bigg\{ \frac{k_{A_1}  (w-\rC) \sqrt{w^2-1} (1 + (\alVL - \alVR)\beVL \rV)}{\sqrt{3} \sqrt{\mSqq }}  +\frac{(k_{A_3} +k_{A_2}  \rC) \big(w^2-1\big)^{3/2} (1 + (\alVL - \alVR)\beVL \rV)}{\sqrt{3} \sqrt{\mSqq }} \notag \\ 
    & -\frac{4 k_{T_1}  \rt (1+w) \sqrt{\mSqq  \big(w^2-1\big)} \alTR  \beTL  \rT}{\sqrt{3} \mSqq }  -\frac{4 k_{T_2}  \rt (w-1) \sqrt{\mSqq  \big(w^2-1\big)} \alTR  \beTL  \rT}{\sqrt{3} \mSqq } \notag \\ 
    & -\frac{4 k_{T_3}  \rt \big(w^2-1\big)^{3/2} \alTR  \beTL  \rT}{\sqrt{3} \sqrt{\mSqq }}\bigg\}\notag \\ 
A_1^{+} & =
 \sin\thtau  e^{+i\phDtau}\bigg\{ -\frac{k_V  \rt \big(w^2-1\big) (1 + (\alVL + \alVR)\beVL \rV)}{2 \sqrt{2} \sqrt{\mSqq }}  +\frac{k_{A_1}  \rt \sqrt{w^2-1} (1 + (\alVL - \alVR)\beVL \rV)}{2 \sqrt{2} \sqrt{\mSqq }} \notag \\ 
    & -\frac{\sqrt{2} k_{T_1}  (1+w) \big((1+\rC) (w-1)+(1-\rC) \sqrt{w^2-1}\big) \alTR  \beTL  \rT}{\sqrt{\mSqq }} \notag \\ 
    & +\frac{\sqrt{2} k_{T_2}  (w-1) \big((1-\rC) (1+w)+(1+\rC) \sqrt{w^2-1}\big) \alTR  \beTL  \rT}{\sqrt{\mSqq }}\bigg\}\notag \\ 
A_2^{+} & =
 \sin^2\frac{\thtau}{2} e^{+i\phDtau}\bigg\{ \frac{k_V  \big(w^2-1\big) (1 + (\alVL + \alVR)\beVL \rV)}{\sqrt{2}}  +\frac{k_{A_1}  \sqrt{w^2-1} (-1 + (\alVR - \alVL)\beVL \rV)}{\sqrt{2}} \notag \\ 
    & +\frac{2 \sqrt{2} k_{T_1}  \rt (1+w) \big((1+\rC) (w-1)+(1-\rC) \sqrt{w^2-1}\big) \alTR  \beTL  \rT}{\mSqq } \notag \\ 
    & -\frac{2 \sqrt{2} k_{T_2}  \rt (w-1) \big((1-\rC) (1+w)+(1+\rC) \sqrt{w^2-1}\big) \alTR  \beTL  \rT}{\mSqq }\bigg\}\notag\,.
\end{align}

\end{widetext}
Finally, the $B\to\Dbar_1^* \ell \nu$ amplitudes are obtained from the $B \to \ov D_1 \ell \nu$ results, with the form factor mapping
\begin{equation}
	f_S \mapsto -g_S\,, \qquad f_{A,V_{i}, T_{i}} \mapsto g_{A,V_{i}, T_{i}}\,,
\end{equation}
where $i=1,2,3$, as follows from the definitions in Eqs.~\eqref{formf121} and~\eqref{formf321}.

\section{$B\to D^{(*)}\ell\bar\nu$}
\label{app:DDsrates}
For completeness, and to have all six $B\to D^{(*,**)}$ rates together
in the same notation, we list here $\d\Gamma/\d w$ for arbitrary charged
lepton mass and weak current for the $B\to D^{(*)}\ell\bar\nu$ modes as well.
We use the form factors defined as in Ref.~\cite{Bernlochner:2017jka}.
For $\Bbar \to D$,
\begin{align}
\ampB{D}{\cbar\,b} & = \sqrt{m_B m_D}\, h_S\, (w+1)\,, \label{eqn:DS} \\*
\ampB{D}{\cbar\g_5 b} & = \ampBs{D}{\cbar \g^\mu\g^5 b} = 0\,, \nn\\
\ampB{D}{\cbar \g_\mu b} & = \sqrt{m_B m_D}\, 
  \big[ h_+(v+v')_\mu + h_-(v-v')_\mu \big], \nn\\*
\ampB{D}{\cbar \sigma_{\mu\nu} b} & = i \sqrt{m_B m_D}\, 
  \big[ h_T\, (v'_\mu v_\nu - v'_\nu v_\mu )\big], \nn
\end{align}
while for the $\Bbar \to D^*$, 
\begin{align}
\ampB{D^*}{\cbar b} & = 0\,, \\*
\ampB{D^*}{\cbar \g_5 b} & = -\sqrt{m_B m_{D^*}}\, h_P\, (\epsilon^* \cdot v)\,, \nn\\
\ampB{D^*}{\cbar \g_\mu b} & = i\sqrt{m_B m_{D^*}}\, h_V\, 
  \varepsilon_{\mu\nu\alpha\beta}\, \epsilon^{*\nu}v'{}^\alpha v^\beta \,,\nn\\
\ampB{D^*}{\cbar \g_\mu \g_5 b} & = \sqrt{m_B m_{D^*}}\, 
  \big[h_{A_1} (w+1)\epsilon^*_\mu \nn\\*
&\quad - h_{A_2}(\epsilon^* \cdot v)v_\mu 
  - h_{A_3}(\epsilon^* \cdot v)v'_\mu \big] , \nn\\
\ampB{D^*}{\cbar \sigma_{\mu\nu} b} & = -\sqrt{m_B m_{D^*}}\,
  \varepsilon_{\mu\nu\alpha\beta} \big[ h_{T_1} \epsilon^{*\alpha}
  (v+v')^\beta \nn\\*
&\quad + h_{T_2} \epsilon^*_{\alpha} (v-v')^\beta 
  + h_{T_3}(\epsilon^*\cdot v) v^\alpha v'{}^\beta \big]. \nn
\end{align}
The common sign convention in $B\to D^{(*)}\ell\bar\nu$ papers is
$\sigma^{\mu\nu} \g^5 \equiv -(i/2)\epsilon^{\mu \nu \rho \sigma}
\sigma_{\rho \sigma}$ such that $\text{Tr}[\g^\mu\g^\nu\g^\sigma\g^\rho\g^5] = +4i
\epsilon^{\mu\nu\rho\sigma}$.  This corresponds to the heavy quark
symmetry relations with signs $h_+ = h_V = h_{A_1} = h_{A_3} = h_S = h_P
= h_T = h_{T_1} = \xi$ (and $h_- =  h_{A_2} = h_{T_2} = h_{T_3} = 0$). This
convention is only used in this appendix, and is the opposite
of that for $B\to D^{(**)}\ell\bar\nu$ used in Refs.~\cite{Leibovich:1997tu,
Leibovich:1997em, Bernlochner:2016bci} and the rest of this paper.

Then we find
\begin{widetext}
\begin{subequations}
\begin{align}
\frac{\d\Gamma_{D}^{\rm (SM)}}{\d w} & = 
  4\, \Gamma_0\, r^3 \sqrt{w^2-1}\, \frac{(\mSqq-\rl\big)^2}{\hat q^6}
  \bigg\{ (w^2-1)\, \mSqq \big[h_+(1+r) - h_-(1-r) \big]^2 \label{DrateSM}\\*
&\quad + \rl \Big[ h_+^2 (w+1) \big[2(w-2r+r^2 w) + \mSqq\big]
  + h_-^2 (w-1) \big[2(w-2r+r^2 w) - \mSqq \big]
  + 4 h_- h_+ (r^2-1) (w^2-1) \Big] \bigg\} \,, \nn \\
\frac{\d\Gamma_D}{\d w} & = 
  \frac{\d\Gamma_D^{\rm (SM)}}{\d w}\,
  \big(1 + \alVLt \beVLt + \alVRt \beVLt\big)^2
  + 4\, \Gamma_0\, r^3 \sqrt{w^2-1}\, \frac{(\mSqq-\rl\big)^2}{\hat q^6}\, 
  \bigg\{ \frac32 \big[(\alSLt + \alSRt) \beSLt\big]^2\, 
    h_S^2\, (w+1)^2\, \hat q^4\label{DrateNP}\\*
&\quad + 3\sqrt{\rl} (\alSLt + \alSRt)\beSLt 
  \big(1 + \alVLt \beVLt + \alVRt \beVLt\big) h_S (1+w)\,
  \mSqq \big[h_- (1+r)(1-w) + h_+(1-r)(1+w)\big] \nn\\*
&\quad + 4 \alTLt \beTLt h_T (w^2-1) \mSqq 
  \Big[ 2 \alTLt \beTLt h_T (\mSqq + 2 \rl)
  + 3\sqrt{\rl} \big(1 + \alVLt \beVLt + \alVRt \beVLt\big)
  \big[h_+(1+r) - h_-(1-r)\big] \Big] \bigg\} \,.\nn
\end{align}
\end{subequations}

The $B\to D^*$ result is
\begin{subequations}
\begin{align}
\frac{\d\Gamma_{D^*}^{\rm (SM)}}{\d w} & = 
  2\Gamma_0 r^3 \sqrt{w^2-1}\, \frac{(\mSqq-\rl\big)^2}{\hat q^6} \bigg\{
  \nn\\*
&\quad 2(w+1)\mSqq \bigg( 2\mSqq \big[h_{A_1}^2(w+1) + h_V^2(w-1)\big]
  + (w+1) \big[h_{A_1}(w-r) - (r\, h_{A_2}+ h_{A_3})(w-1) \big]^2
  \bigg) \label{DsrateSM}\\
&\quad + \rl (w+1) \bigg(
  \big[h_{A_1}^2 (w+1)+h_{A_3}^2 (w-1)\big] [4(w-r)^2-\mSqq]
  + 2 h_{A_1}(w^2-1) \big[h_{A_2} (r^2+2 r w-3)+4 h_{A_3} (r-w)\big] \nn\\
&\qquad +(w-1) \Big[ \big(2 h_V^2 - h_{A_2}^2\big) \mSqq 
+ h_{A_2}^2 4(r w-1)^2
+2 h_{A_2} h_{A_3} \big[3 w + 3 r^2 w-2 r (w^2+2)\big] \Big]
 \bigg) \bigg\}\,.\nn\\
\frac{\d\Gamma_{D^*}}{\d w} & = 
  \frac{\d\Gamma_{D^*}^{\rm (SM)}}{\d w}\,
  \big(1+ \alVLt \beVLt-\alVRt \beVLt\big)^2
  + 2\Gamma_0 r^3 \sqrt{w^2-1}\, \frac{(\mSqq-\rl\big)^2}{\hat q^6}\, 
  \bigg\{ 3 \big[(\alSLt-\alSRt) \beSLt\big]^2 h_P^2 (w^2-1) \hat q^4 \label{DsrateNP}\\*
&\quad - 6 (\alSLt - \alSRt) \beSLt
  \big(1 + \alVLt \beVLt - \alVRt \beVLt\big)
  h_P (w^2-1) \mSqq \sqrt{\rl}\,
  \big[h_{A_1}(1+w) - h_{A_2}(1-r w) - h_{A_3} (w-r) \big] \nn\\*
&\quad + 16 (\alTLt \beTLt)^2 (\mSqq+2 \rl) 
  \bigg(h_{T_1}^2 (w+1) \big[\mSqq(5w+1) + 8r(w^2-1)] 
  + h_{T_2}^2 (w-1) [\mSqq(5w-1) + 8r(w^2-1)] \nn\\*
&\qquad + h_{T_3}^2 \mSqq (w^2-1)^2
  - 2 h_{T_1} h_{T_2} (w^2-1) (3 +2rw-5r^2)
  - 2 h_{T_3} \big[h_{T_1}(w+1)+h_{T_2}(w-1)\big] \mSqq(w^2-1) \bigg) \nn\\*
&\quad + 24 \alTLt \beTLt \sqrt{\rl}\, \mSqq (w+1) \bigg(
  \big(1 + \alVLt \beVLt + \alVRt \beVLt\big)\,
  2 h_V \big[h_{T_1}(1+r) - h_{T_2}(1-r)\big] (w-1) \nn\\*
&\qquad + \big(1 + \alVLt \beVLt - \alVRt \beVLt\big)
  \Big[ h_{A_1} \big[ h_{T_2} (2-w+3r)(w-1) - h_{T_1} (2+w-3r)(w+1)
  + h_{T_3} (w-r)(w^2-1) \big] \nn\\*
&\qquad + (h_{A_2} r + h_{A_3})
  \big[h_{T_1}(w+1) + h_{T_2}(w-1) - h_{T_3}(w^2-1)\big] (w-1) \Big]\bigg) \nn\\*
&\quad + 8 \alVRt \beVLt (1+\alVLt \beVLt)\, h_V^2\,
  \mSqq (2\mSqq+\rl) (w^2-1) \bigg\} \,. \nn
\end{align}
\end{subequations}

\twocolumngrid
\end{widetext}

\bibliographystyle{apsrev4-1}

\begin{thebibliography}{38}%
\makeatletter
\providecommand \@ifxundefined [1]{%
 \@ifx{#1\undefined}
}%
\providecommand \@ifnum [1]{%
 \ifnum #1\expandafter \@firstoftwo
 \else \expandafter \@secondoftwo
 \fi
}%
\providecommand \@ifx [1]{%
 \ifx #1\expandafter \@firstoftwo
 \else \expandafter \@secondoftwo
 \fi
}%
\providecommand \natexlab [1]{#1}%
\providecommand \enquote  [1]{``#1''}%
\providecommand \bibnamefont  [1]{#1}%
\providecommand \bibfnamefont [1]{#1}%
\providecommand \citenamefont [1]{#1}%
\providecommand \href@noop [0]{\@secondoftwo}%
\providecommand \href [0]{\begingroup \@sanitize@url \@href}%
\providecommand \@href[1]{\@@startlink{#1}\@@href}%
\providecommand \@@href[1]{\endgroup#1\@@endlink}%
\providecommand \@sanitize@url [0]{\catcode `\\12\catcode `\$12\catcode
  `\&12\catcode `\#12\catcode `\^12\catcode `\_12\catcode `\%12\relax}%
\providecommand \@@startlink[1]{}%
\providecommand \@@endlink[0]{}%
\providecommand \url  [0]{\begingroup\@sanitize@url \@url }%
\providecommand \@url [1]{\endgroup\@href {#1}{\urlprefix }}%
\providecommand \urlprefix  [0]{URL }%
\providecommand \Eprint [0]{\href }%
\providecommand \doibase [0]{http://dx.doi.org/}%
\providecommand \selectlanguage [0]{\@gobble}%
\providecommand \bibinfo  [0]{\@secondoftwo}%
\providecommand \bibfield  [0]{\@secondoftwo}%
\providecommand \translation [1]{[#1]}%
\providecommand \BibitemOpen [0]{}%
\providecommand \bibitemStop [0]{}%
\providecommand \bibitemNoStop [0]{.\EOS\space}%
\providecommand \EOS [0]{\spacefactor3000\relax}%
\providecommand \BibitemShut  [1]{\csname bibitem#1\endcsname}%
\let\auto@bib@innerbib\@empty
\bibitem [{\citenamefont {Amhis}\ \emph {et~al.}(2016)\citenamefont {Amhis}
  \emph {et~al.}}]{HFAG}%
  \BibitemOpen
  \bibfield  {author} {\bibinfo {author} {\bibfnamefont {Y.}~\bibnamefont
  {Amhis}} \emph {et~al.} (\bibinfo {collaboration} {Heavy Flavor Averaging
  Group}),\ }\href@noop {} {\  (\bibinfo {year} {2016})},\ \bibinfo {note} {and
  updates at \url{http://www.slac.stanford.edu/xorg/hfag/}},\ \Eprint
  {http://arxiv.org/abs/1612.07233} {arXiv:1612.07233 [hep-ex]} \BibitemShut
  {NoStop}%
\bibitem [{\citenamefont {Bernlochner}\ \emph
  {et~al.}(2017{\natexlab{a}})\citenamefont {Bernlochner}, \citenamefont
  {Ligeti}, \citenamefont {Papucci},\ and\ \citenamefont
  {Robinson}}]{Bernlochner:2017jka}%
  \BibitemOpen
  \bibfield  {author} {\bibinfo {author} {\bibfnamefont {F.~U.}\ \bibnamefont
  {Bernlochner}}, \bibinfo {author} {\bibfnamefont {Z.}~\bibnamefont {Ligeti}},
  \bibinfo {author} {\bibfnamefont {M.}~\bibnamefont {Papucci}}, \ and\
  \bibinfo {author} {\bibfnamefont {D.~J.}\ \bibnamefont {Robinson}},\ }\href
  {\doibase 10.1103/PhysRevD.95.115008} {\bibfield  {journal} {\bibinfo
  {journal} {Phys. Rev.}\ }\textbf {\bibinfo {volume} {D95}},\ \bibinfo {pages}
  {115008} (\bibinfo {year} {2017}{\natexlab{a}})},\ \Eprint
  {http://arxiv.org/abs/1703.05330} {arXiv:1703.05330 [hep-ph]} \BibitemShut
  {NoStop}%
\bibitem [{\citenamefont {Bigi}\ \emph {et~al.}(2017)\citenamefont {Bigi},
  \citenamefont {Gambino},\ and\ \citenamefont {Schacht}}]{Bigi:2017njr}%
  \BibitemOpen
  \bibfield  {author} {\bibinfo {author} {\bibfnamefont {D.}~\bibnamefont
  {Bigi}}, \bibinfo {author} {\bibfnamefont {P.}~\bibnamefont {Gambino}}, \
  and\ \bibinfo {author} {\bibfnamefont {S.}~\bibnamefont {Schacht}},\ }\href
  {\doibase 10.1016/j.physletb.2017.04.022} {\bibfield  {journal} {\bibinfo
  {journal} {Phys. Lett.}\ }\textbf {\bibinfo {volume} {B769}},\ \bibinfo
  {pages} {441} (\bibinfo {year} {2017})},\ \Eprint
  {http://arxiv.org/abs/1703.06124} {arXiv:1703.06124 [hep-ph]} \BibitemShut
  {NoStop}%
\bibitem [{\citenamefont {Grinstein}\ and\ \citenamefont
  {Kobach}(2017)}]{Grinstein:2017nlq}%
  \BibitemOpen
  \bibfield  {author} {\bibinfo {author} {\bibfnamefont {B.}~\bibnamefont
  {Grinstein}}\ and\ \bibinfo {author} {\bibfnamefont {A.}~\bibnamefont
  {Kobach}},\ }\href {\doibase 10.1016/j.physletb.2017.05.078} {\bibfield
  {journal} {\bibinfo  {journal} {Phys. Lett.}\ }\textbf {\bibinfo {volume}
  {B771}},\ \bibinfo {pages} {359} (\bibinfo {year} {2017})},\ \Eprint
  {http://arxiv.org/abs/1703.08170} {arXiv:1703.08170 [hep-ph]} \BibitemShut
  {NoStop}%
\bibitem [{\citenamefont {Bernlochner}\ \emph
  {et~al.}(2017{\natexlab{b}})\citenamefont {Bernlochner}, \citenamefont
  {Ligeti}, \citenamefont {Papucci},\ and\ \citenamefont
  {Robinson}}]{Bernlochner:2017xyx}%
  \BibitemOpen
  \bibfield  {author} {\bibinfo {author} {\bibfnamefont {F.~U.}\ \bibnamefont
  {Bernlochner}}, \bibinfo {author} {\bibfnamefont {Z.}~\bibnamefont {Ligeti}},
  \bibinfo {author} {\bibfnamefont {M.}~\bibnamefont {Papucci}}, \ and\
  \bibinfo {author} {\bibfnamefont {D.~J.}\ \bibnamefont {Robinson}},\
  }\href@noop {} {\  (\bibinfo {year} {2017}{\natexlab{b}})},\ \Eprint
  {http://arxiv.org/abs/1708.07134} {arXiv:1708.07134 [hep-ph]} \BibitemShut
  {NoStop}%
\bibitem [{\citenamefont {Bailey}\ \emph {et~al.}(2015)\citenamefont {Bailey}
  \emph {et~al.}}]{Lattice:2015rga}%
  \BibitemOpen
  \bibfield  {author} {\bibinfo {author} {\bibfnamefont {J.~A.}\ \bibnamefont
  {Bailey}} \emph {et~al.} (\bibinfo {collaboration} {Fermilab Lattice and MILC
  Collaborations}),\ }\href {\doibase 10.1103/PhysRevD.92.034506} {\bibfield
  {journal} {\bibinfo  {journal} {Phys. Rev.}\ }\textbf {\bibinfo {volume}
  {D92}},\ \bibinfo {pages} {034506} (\bibinfo {year} {2015})},\ \Eprint
  {http://arxiv.org/abs/1503.07237} {arXiv:1503.07237 [hep-lat]} \BibitemShut
  {NoStop}%
\bibitem [{\citenamefont {Vaquero}\ \emph {et~al.}(2017)\citenamefont {Vaquero}
  \emph {et~al.}}]{BDsLatticeAllw}%
  \BibitemOpen
  \bibfield  {author} {\bibinfo {author} {\bibfnamefont {A.}~\bibnamefont
  {Vaquero}} \emph {et~al.} (\bibinfo {collaboration} {Fermilab Lattice and
  MILC Collaborations}),\ }\href@noop {} {\bibfield  {journal} {\bibinfo
  {journal} {Talk at the Lattice 2017 Conference,}\ } (\bibinfo {year}
  {2017})},\ \bibinfo {note}
  {\url{https://makondo.ugr.es/event/0/session/92/contribution/120/material/slides/0.pdf}}\BibitemShut
  {NoStop}%
\bibitem [{\citenamefont {Jaiswal}\ \emph {et~al.}(2017)\citenamefont
  {Jaiswal}, \citenamefont {Nandi},\ and\ \citenamefont
  {Patra}}]{Jaiswal:2017rve}%
  \BibitemOpen
  \bibfield  {author} {\bibinfo {author} {\bibfnamefont {S.}~\bibnamefont
  {Jaiswal}}, \bibinfo {author} {\bibfnamefont {S.}~\bibnamefont {Nandi}}, \
  and\ \bibinfo {author} {\bibfnamefont {S.~K.}\ \bibnamefont {Patra}},\
  }\href@noop {} {\  (\bibinfo {year} {2017})},\ \Eprint
  {http://arxiv.org/abs/1707.09977} {arXiv:1707.09977 [hep-ph]} \BibitemShut
  {NoStop}%
\bibitem [{\citenamefont {Isgur}\ and\ \citenamefont
  {Wise}(1989)}]{Isgur:1989vq}%
  \BibitemOpen
  \bibfield  {author} {\bibinfo {author} {\bibfnamefont {N.}~\bibnamefont
  {Isgur}}\ and\ \bibinfo {author} {\bibfnamefont {M.~B.}\ \bibnamefont
  {Wise}},\ }\href {\doibase 10.1016/0370-2693(89)90566-2} {\bibfield
  {journal} {\bibinfo  {journal} {Phys. Lett.}\ }\textbf {\bibinfo {volume}
  {B232}},\ \bibinfo {pages} {113} (\bibinfo {year} {1989})}\BibitemShut
  {NoStop}%
\bibitem [{\citenamefont {Isgur}\ and\ \citenamefont
  {Wise}(1990)}]{Isgur:1989ed}%
  \BibitemOpen
  \bibfield  {author} {\bibinfo {author} {\bibfnamefont {N.}~\bibnamefont
  {Isgur}}\ and\ \bibinfo {author} {\bibfnamefont {M.~B.}\ \bibnamefont
  {Wise}},\ }\href {\doibase 10.1016/0370-2693(90)91219-2} {\bibfield
  {journal} {\bibinfo  {journal} {Phys. Lett.}\ }\textbf {\bibinfo {volume}
  {B237}},\ \bibinfo {pages} {527} (\bibinfo {year} {1990})}\BibitemShut
  {NoStop}%
\bibitem [{\citenamefont {Isgur}\ and\ \citenamefont
  {Wise}(1991{\natexlab{a}})}]{Isgur:1991wq}%
  \BibitemOpen
  \bibfield  {author} {\bibinfo {author} {\bibfnamefont {N.}~\bibnamefont
  {Isgur}}\ and\ \bibinfo {author} {\bibfnamefont {M.~B.}\ \bibnamefont
  {Wise}},\ }\href {\doibase 10.1103/PhysRevLett.66.1130} {\bibfield  {journal}
  {\bibinfo  {journal} {Phys. Rev. Lett.}\ }\textbf {\bibinfo {volume} {66}},\
  \bibinfo {pages} {1130} (\bibinfo {year} {1991}{\natexlab{a}})}\BibitemShut
  {NoStop}%
\bibitem [{\citenamefont {Bernlochner}\ and\ \citenamefont
  {Ligeti}(2017)}]{Bernlochner:2016bci}%
  \BibitemOpen
  \bibfield  {author} {\bibinfo {author} {\bibfnamefont {F.~U.}\ \bibnamefont
  {Bernlochner}}\ and\ \bibinfo {author} {\bibfnamefont {Z.}~\bibnamefont
  {Ligeti}},\ }\href {\doibase 10.1103/PhysRevD.95.014022} {\bibfield
  {journal} {\bibinfo  {journal} {Phys. Rev.}\ }\textbf {\bibinfo {volume}
  {D95}},\ \bibinfo {pages} {014022} (\bibinfo {year} {2017})},\ \Eprint
  {http://arxiv.org/abs/1606.09300} {arXiv:1606.09300 [hep-ph]} \BibitemShut
  {NoStop}%
\bibitem [{\citenamefont {Leibovich}\ \emph {et~al.}(1997)\citenamefont
  {Leibovich}, \citenamefont {Ligeti}, \citenamefont {Stewart},\ and\
  \citenamefont {Wise}}]{Leibovich:1997tu}%
  \BibitemOpen
  \bibfield  {author} {\bibinfo {author} {\bibfnamefont {A.~K.}\ \bibnamefont
  {Leibovich}}, \bibinfo {author} {\bibfnamefont {Z.}~\bibnamefont {Ligeti}},
  \bibinfo {author} {\bibfnamefont {I.~W.}\ \bibnamefont {Stewart}}, \ and\
  \bibinfo {author} {\bibfnamefont {M.~B.}\ \bibnamefont {Wise}},\ }\href
  {\doibase 10.1103/PhysRevLett.78.3995} {\bibfield  {journal} {\bibinfo
  {journal} {Phys. Rev. Lett.}\ }\textbf {\bibinfo {volume} {78}},\ \bibinfo
  {pages} {3995} (\bibinfo {year} {1997})},\ \Eprint
  {http://arxiv.org/abs/hep-ph/9703213} {arXiv:hep-ph/9703213 [hep-ph]}
  \BibitemShut {NoStop}%
\bibitem [{\citenamefont {Leibovich}\ \emph {et~al.}(1998)\citenamefont
  {Leibovich}, \citenamefont {Ligeti}, \citenamefont {Stewart},\ and\
  \citenamefont {Wise}}]{Leibovich:1997em}%
  \BibitemOpen
  \bibfield  {author} {\bibinfo {author} {\bibfnamefont {A.~K.}\ \bibnamefont
  {Leibovich}}, \bibinfo {author} {\bibfnamefont {Z.}~\bibnamefont {Ligeti}},
  \bibinfo {author} {\bibfnamefont {I.~W.}\ \bibnamefont {Stewart}}, \ and\
  \bibinfo {author} {\bibfnamefont {M.~B.}\ \bibnamefont {Wise}},\ }\href
  {\doibase 10.1103/PhysRevD.57.308} {\bibfield  {journal} {\bibinfo  {journal}
  {Phys. Rev.}\ }\textbf {\bibinfo {volume} {D57}},\ \bibinfo {pages} {308}
  (\bibinfo {year} {1998})},\ \Eprint {http://arxiv.org/abs/hep-ph/9705467}
  {arXiv:hep-ph/9705467 [hep-ph]} \BibitemShut {NoStop}%
\bibitem [{\citenamefont {Georgi}(1990)}]{Georgi:1990um}%
  \BibitemOpen
  \bibfield  {author} {\bibinfo {author} {\bibfnamefont {H.}~\bibnamefont
  {Georgi}},\ }\href {\doibase 10.1016/0370-2693(90)91128-X} {\bibfield
  {journal} {\bibinfo  {journal} {Phys. Lett.}\ }\textbf {\bibinfo {volume}
  {B240}},\ \bibinfo {pages} {447} (\bibinfo {year} {1990})}\BibitemShut
  {NoStop}%
\bibitem [{\citenamefont {Eichten}\ and\ \citenamefont
  {Hill}(1990{\natexlab{a}})}]{Eichten:1989zv}%
  \BibitemOpen
  \bibfield  {author} {\bibinfo {author} {\bibfnamefont {E.}~\bibnamefont
  {Eichten}}\ and\ \bibinfo {author} {\bibfnamefont {B.~R.}\ \bibnamefont
  {Hill}},\ }\href {\doibase 10.1016/0370-2693(90)92049-O} {\bibfield
  {journal} {\bibinfo  {journal} {Phys. Lett.}\ }\textbf {\bibinfo {volume}
  {B234}},\ \bibinfo {pages} {511} (\bibinfo {year}
  {1990}{\natexlab{a}})}\BibitemShut {NoStop}%
\bibitem [{\citenamefont {Isgur}\ and\ \citenamefont
  {Wise}(1991{\natexlab{b}})}]{Isgur:1990jf}%
  \BibitemOpen
  \bibfield  {author} {\bibinfo {author} {\bibfnamefont {N.}~\bibnamefont
  {Isgur}}\ and\ \bibinfo {author} {\bibfnamefont {M.~B.}\ \bibnamefont
  {Wise}},\ }\href {\doibase 10.1103/PhysRevD.43.819} {\bibfield  {journal}
  {\bibinfo  {journal} {Phys. Rev.}\ }\textbf {\bibinfo {volume} {D43}},\
  \bibinfo {pages} {819} (\bibinfo {year} {1991}{\natexlab{b}})}\BibitemShut
  {NoStop}%
\bibitem [{\citenamefont {Bjorken}(1990{\natexlab{a}})}]{Bjorken:1990hs}%
  \BibitemOpen
  \bibfield  {author} {\bibinfo {author} {\bibfnamefont {J.~D.}\ \bibnamefont
  {Bjorken}},\ }\bibfield  {booktitle} {\emph {\bibinfo {booktitle}
  {{Proceedings, 4th Les Rencontres de Physique de la Vallee d'Aoste: Results
  and Perspectives in Particle Physics: La Thuile, France, March 18-24,
  1990}}},\ }\href@noop {} {\bibfield  {journal} {\bibinfo  {journal} {Conf.
  Proc.}\ }\textbf {\bibinfo {volume} {C900318}},\ \bibinfo {pages} {583}
  (\bibinfo {year} {1990}{\natexlab{a}})}\BibitemShut {NoStop}%
\bibitem [{\citenamefont {Patrignani}\ \emph {et~al.}(2016)\citenamefont
  {Patrignani} \emph {et~al.}}]{PDG}%
  \BibitemOpen
  \bibfield  {author} {\bibinfo {author} {\bibfnamefont {C.}~\bibnamefont
  {Patrignani}} \emph {et~al.} (\bibinfo {collaboration} {Particle Data
  Group}),\ }\href {\doibase 10.1088/1674-1137/40/10/100001} {\bibfield
  {journal} {\bibinfo  {journal} {Chin. Phys.}\ }\textbf {\bibinfo {volume}
  {C40}},\ \bibinfo {pages} {100001} (\bibinfo {year} {2016})}\BibitemShut
  {NoStop}%
\bibitem [{\citenamefont {Falk}\ \emph {et~al.}(1990)\citenamefont {Falk},
  \citenamefont {Georgi}, \citenamefont {Grinstein},\ and\ \citenamefont
  {Wise}}]{Falk:1990yz}%
  \BibitemOpen
  \bibfield  {author} {\bibinfo {author} {\bibfnamefont {A.~F.}\ \bibnamefont
  {Falk}}, \bibinfo {author} {\bibfnamefont {H.}~\bibnamefont {Georgi}},
  \bibinfo {author} {\bibfnamefont {B.}~\bibnamefont {Grinstein}}, \ and\
  \bibinfo {author} {\bibfnamefont {M.~B.}\ \bibnamefont {Wise}},\ }\href
  {\doibase 10.1016/0550-3213(90)90591-Z} {\bibfield  {journal} {\bibinfo
  {journal} {Nucl. Phys.}\ }\textbf {\bibinfo {volume} {B343}},\ \bibinfo
  {pages} {1} (\bibinfo {year} {1990})}\BibitemShut {NoStop}%
\bibitem [{\citenamefont {Bjorken}(1990{\natexlab{b}})}]{Bjorken:1990rr}%
  \BibitemOpen
  \bibfield  {author} {\bibinfo {author} {\bibfnamefont {J.~D.}\ \bibnamefont
  {Bjorken}},\ }in\ \href
  {http://www-public.slac.stanford.edu/sciDoc/docMeta.aspx?slacPubNumber=SLAC-PUB-5389}
  {\emph {\bibinfo {booktitle} {{Gauge bosons and heavy quarks: Proceedings,
  18th SLAC Summer Institute on Particle Physics (SSI 90), Jul 16-27, 1990}}}}\
  (\bibinfo {year} {1990})\ pp.\ \bibinfo {pages} {0167--198}\BibitemShut
  {NoStop}%
\bibitem [{\citenamefont {Falk}(1992)}]{Falk:1991nq}%
  \BibitemOpen
  \bibfield  {author} {\bibinfo {author} {\bibfnamefont {A.~F.}\ \bibnamefont
  {Falk}},\ }\href {\doibase 10.1016/0550-3213(92)90004-U} {\bibfield
  {journal} {\bibinfo  {journal} {Nucl. Phys.}\ }\textbf {\bibinfo {volume}
  {B378}},\ \bibinfo {pages} {79} (\bibinfo {year} {1992})}\BibitemShut
  {NoStop}%
\bibitem [{\citenamefont {Manohar}\ and\ \citenamefont
  {Wise}(2000)}]{Manohar:2000dt}%
  \BibitemOpen
  \bibfield  {author} {\bibinfo {author} {\bibfnamefont {A.~V.}\ \bibnamefont
  {Manohar}}\ and\ \bibinfo {author} {\bibfnamefont {M.~B.}\ \bibnamefont
  {Wise}},\ }\href@noop {} {\bibfield  {journal} {\bibinfo  {journal} {Camb.
  Monogr. Part. Phys. Nucl. Phys. Cosmol.}\ }\textbf {\bibinfo {volume} {10}},\
  \bibinfo {pages} {1} (\bibinfo {year} {2000})}\BibitemShut {NoStop}%
\bibitem [{\citenamefont {Falk}\ and\ \citenamefont
  {Grinstein}(1990)}]{Falk:1990cz}%
  \BibitemOpen
  \bibfield  {author} {\bibinfo {author} {\bibfnamefont {A.~F.}\ \bibnamefont
  {Falk}}\ and\ \bibinfo {author} {\bibfnamefont {B.}~\bibnamefont
  {Grinstein}},\ }\href {\doibase 10.1016/0370-2693(90)91262-A} {\bibfield
  {journal} {\bibinfo  {journal} {Phys. Lett.}\ }\textbf {\bibinfo {volume}
  {B249}},\ \bibinfo {pages} {314} (\bibinfo {year} {1990})}\BibitemShut
  {NoStop}%
\bibitem [{\citenamefont {Neubert}(1992)}]{Neubert:1992qq}%
  \BibitemOpen
  \bibfield  {author} {\bibinfo {author} {\bibfnamefont {M.}~\bibnamefont
  {Neubert}},\ }\href {\doibase 10.1016/0550-3213(92)90233-2} {\bibfield
  {journal} {\bibinfo  {journal} {Nucl. Phys.}\ }\textbf {\bibinfo {volume}
  {B371}},\ \bibinfo {pages} {149} (\bibinfo {year} {1992})}\BibitemShut
  {NoStop}%
\bibitem [{\citenamefont {Falk}\ and\ \citenamefont
  {Neubert}(1993)}]{Falk:1992wt}%
  \BibitemOpen
  \bibfield  {author} {\bibinfo {author} {\bibfnamefont {A.~F.}\ \bibnamefont
  {Falk}}\ and\ \bibinfo {author} {\bibfnamefont {M.}~\bibnamefont {Neubert}},\
  }\href {\doibase 10.1103/PhysRevD.47.2965} {\bibfield  {journal} {\bibinfo
  {journal} {Phys. Rev.}\ }\textbf {\bibinfo {volume} {D47}},\ \bibinfo {pages}
  {2965} (\bibinfo {year} {1993})},\ \Eprint
  {http://arxiv.org/abs/hep-ph/9209268} {arXiv:hep-ph/9209268 [hep-ph]}
  \BibitemShut {NoStop}%
\bibitem [{\citenamefont {Eichten}\ and\ \citenamefont
  {Hill}(1990{\natexlab{b}})}]{Eichten:1990vp}%
  \BibitemOpen
  \bibfield  {author} {\bibinfo {author} {\bibfnamefont {E.}~\bibnamefont
  {Eichten}}\ and\ \bibinfo {author} {\bibfnamefont {B.~R.}\ \bibnamefont
  {Hill}},\ }\href {\doibase 10.1016/0370-2693(90)91408-4} {\bibfield
  {journal} {\bibinfo  {journal} {Phys. Lett.}\ }\textbf {\bibinfo {volume}
  {B243}},\ \bibinfo {pages} {427} (\bibinfo {year}
  {1990}{\natexlab{b}})}\BibitemShut {NoStop}%
\bibitem [{\citenamefont {Luke}(1990)}]{Luke:1990eg}%
  \BibitemOpen
  \bibfield  {author} {\bibinfo {author} {\bibfnamefont {M.~E.}\ \bibnamefont
  {Luke}},\ }\href {\doibase 10.1016/0370-2693(90)90568-Q} {\bibfield
  {journal} {\bibinfo  {journal} {Phys. Lett.}\ }\textbf {\bibinfo {volume}
  {B252}},\ \bibinfo {pages} {447} (\bibinfo {year} {1990})}\BibitemShut
  {NoStop}%
\bibitem [{\citenamefont {Falk}\ \emph {et~al.}(1991)\citenamefont {Falk},
  \citenamefont {Grinstein},\ and\ \citenamefont {Luke}}]{Falk:1990pz}%
  \BibitemOpen
  \bibfield  {author} {\bibinfo {author} {\bibfnamefont {A.~F.}\ \bibnamefont
  {Falk}}, \bibinfo {author} {\bibfnamefont {B.}~\bibnamefont {Grinstein}}, \
  and\ \bibinfo {author} {\bibfnamefont {M.~E.}\ \bibnamefont {Luke}},\ }\href
  {\doibase 10.1016/0550-3213(91)90464-9} {\bibfield  {journal} {\bibinfo
  {journal} {Nucl. Phys.}\ }\textbf {\bibinfo {volume} {B357}},\ \bibinfo
  {pages} {185} (\bibinfo {year} {1991})}\BibitemShut {NoStop}%
\bibitem [{\citenamefont {Manohar}(1997)}]{Manohar:1996cq}%
  \BibitemOpen
  \bibfield  {author} {\bibinfo {author} {\bibfnamefont {A.~V.}\ \bibnamefont
  {Manohar}},\ }\bibfield  {booktitle} {\emph {\bibinfo {booktitle}
  {{Perturbative and nonperturbative aspects of quantum field theory.
  Proceedings, 35. Internationale Universit{\"a}tswochen f{\"u}r Kern- und
  Teilchenphysik: Schladming, Austria, March 2-9, 1996}}},\ }\href {\doibase
  10.1007/BFb0104294} {\bibfield  {journal} {\bibinfo  {journal} {Lect. Notes
  Phys.}\ }\textbf {\bibinfo {volume} {479}},\ \bibinfo {pages} {311} (\bibinfo
  {year} {1997})},\ \Eprint {http://arxiv.org/abs/hep-ph/9606222}
  {arXiv:hep-ph/9606222 [hep-ph]} \BibitemShut {NoStop}%
\bibitem [{\citenamefont {Ligeti}\ \emph {et~al.}(2017)\citenamefont {Ligeti},
  \citenamefont {Papucci},\ and\ \citenamefont {Robinson}}]{Ligeti:2016npd}%
  \BibitemOpen
  \bibfield  {author} {\bibinfo {author} {\bibfnamefont {Z.}~\bibnamefont
  {Ligeti}}, \bibinfo {author} {\bibfnamefont {M.}~\bibnamefont {Papucci}}, \
  and\ \bibinfo {author} {\bibfnamefont {D.~J.}\ \bibnamefont {Robinson}},\
  }\href {\doibase 10.1007/JHEP01(2017)083} {\bibfield  {journal} {\bibinfo
  {journal} {JHEP}\ }\textbf {\bibinfo {volume} {01}},\ \bibinfo {pages} {083}
  (\bibinfo {year} {2017})},\ \Eprint {http://arxiv.org/abs/1610.02045}
  {arXiv:1610.02045 [hep-ph]} \BibitemShut {NoStop}%
\bibitem [{\citenamefont {Biancofiore}\ \emph {et~al.}(2013)\citenamefont
  {Biancofiore}, \citenamefont {Colangelo},\ and\ \citenamefont
  {De~Fazio}}]{Biancofiore:2013ki}%
  \BibitemOpen
  \bibfield  {author} {\bibinfo {author} {\bibfnamefont {P.}~\bibnamefont
  {Biancofiore}}, \bibinfo {author} {\bibfnamefont {P.}~\bibnamefont
  {Colangelo}}, \ and\ \bibinfo {author} {\bibfnamefont {F.}~\bibnamefont
  {De~Fazio}},\ }\href {\doibase 10.1103/PhysRevD.87.074010} {\bibfield
  {journal} {\bibinfo  {journal} {Phys. Rev.}\ }\textbf {\bibinfo {volume}
  {D87}},\ \bibinfo {pages} {074010} (\bibinfo {year} {2013})},\ \Eprint
  {http://arxiv.org/abs/1302.1042} {arXiv:1302.1042 [hep-ph]} \BibitemShut
  {NoStop}%
\bibitem [{\citenamefont {Duraisamy}\ \emph {et~al.}(2014)\citenamefont
  {Duraisamy}, \citenamefont {Sharma},\ and\ \citenamefont
  {Datta}}]{Duraisamy:2014sna}%
  \BibitemOpen
  \bibfield  {author} {\bibinfo {author} {\bibfnamefont {M.}~\bibnamefont
  {Duraisamy}}, \bibinfo {author} {\bibfnamefont {P.}~\bibnamefont {Sharma}}, \
  and\ \bibinfo {author} {\bibfnamefont {A.}~\bibnamefont {Datta}},\ }\href
  {\doibase 10.1103/PhysRevD.90.074013} {\bibfield  {journal} {\bibinfo
  {journal} {Phys. Rev.}\ }\textbf {\bibinfo {volume} {D90}},\ \bibinfo {pages}
  {074013} (\bibinfo {year} {2014})},\ \Eprint {http://arxiv.org/abs/1405.3719}
  {arXiv:1405.3719 [hep-ph]} \BibitemShut {NoStop}%
\bibitem [{\citenamefont {Liventsev}\ \emph {et~al.}(2008)\citenamefont
  {Liventsev} \emph {et~al.}}]{Liventsev:2007rb}%
  \BibitemOpen
  \bibfield  {author} {\bibinfo {author} {\bibfnamefont {D.}~\bibnamefont
  {Liventsev}} \emph {et~al.} (\bibinfo {collaboration} {Belle}),\ }\href
  {\doibase 10.1103/PhysRevD.77.091503} {\bibfield  {journal} {\bibinfo
  {journal} {Phys. Rev.}\ }\textbf {\bibinfo {volume} {D77}},\ \bibinfo {pages}
  {091503} (\bibinfo {year} {2008})},\ \Eprint {http://arxiv.org/abs/0711.3252}
  {arXiv:0711.3252 [hep-ex]} \BibitemShut {NoStop}%
\bibitem [{\citenamefont {Freytsis}\ \emph {et~al.}(2015)\citenamefont
  {Freytsis}, \citenamefont {Ligeti},\ and\ \citenamefont
  {Ruderman}}]{Freytsis:2015qca}%
  \BibitemOpen
  \bibfield  {author} {\bibinfo {author} {\bibfnamefont {M.}~\bibnamefont
  {Freytsis}}, \bibinfo {author} {\bibfnamefont {Z.}~\bibnamefont {Ligeti}}, \
  and\ \bibinfo {author} {\bibfnamefont {J.~T.}\ \bibnamefont {Ruderman}},\
  }\href {\doibase 10.1103/PhysRevD.92.054018} {\bibfield  {journal} {\bibinfo
  {journal} {Phys. Rev.}\ }\textbf {\bibinfo {volume} {D92}},\ \bibinfo {pages}
  {054018} (\bibinfo {year} {2015})},\ \Eprint
  {http://arxiv.org/abs/1506.08896} {arXiv:1506.08896 [hep-ph]} \BibitemShut
  {NoStop}%
\bibitem [{\citenamefont {Huschle}\ \emph {et~al.}(2015)\citenamefont {Huschle}
  \emph {et~al.}}]{Huschle:2015rga}%
  \BibitemOpen
  \bibfield  {author} {\bibinfo {author} {\bibfnamefont {M.}~\bibnamefont
  {Huschle}} \emph {et~al.} (\bibinfo {collaboration} {Belle Collaboration}),\
  }\href {\doibase 10.1103/PhysRevD.92.072014} {\bibfield  {journal} {\bibinfo
  {journal} {Phys. Rev.}\ }\textbf {\bibinfo {volume} {D92}},\ \bibinfo {pages}
  {072014} (\bibinfo {year} {2015})},\ \Eprint
  {http://arxiv.org/abs/1507.03233} {arXiv:1507.03233 [hep-ex]} \BibitemShut
  {NoStop}%
\bibitem [{\citenamefont {Lees}\ \emph {et~al.}(2013)\citenamefont {Lees} \emph
  {et~al.}}]{Lees:2013uzd}%
  \BibitemOpen
  \bibfield  {author} {\bibinfo {author} {\bibfnamefont {J.~P.}\ \bibnamefont
  {Lees}} \emph {et~al.} (\bibinfo {collaboration} {BaBar Collaboration}),\
  }\href {\doibase 10.1103/PhysRevD.88.072012} {\bibfield  {journal} {\bibinfo
  {journal} {Phys. Rev.}\ }\textbf {\bibinfo {volume} {D88}},\ \bibinfo {pages}
  {072012} (\bibinfo {year} {2013})},\ \Eprint {http://arxiv.org/abs/1303.0571}
  {arXiv:1303.0571 [hep-ex]} \BibitemShut {NoStop}%
\bibitem [{\citenamefont {Duell}\ \emph {et~al.}(2017)\citenamefont {Duell},
  \citenamefont {Bernlochner}, \citenamefont {Ligeti}, \citenamefont
  {Papucci},\ and\ \citenamefont {Robinson}}]{Duell:2016maj}%
  \BibitemOpen
  \bibfield  {author} {\bibinfo {author} {\bibfnamefont {S.}~\bibnamefont
  {Duell}}, \bibinfo {author} {\bibfnamefont {F.}~\bibnamefont {Bernlochner}},
  \bibinfo {author} {\bibfnamefont {Z.}~\bibnamefont {Ligeti}}, \bibinfo
  {author} {\bibfnamefont {M.}~\bibnamefont {Papucci}}, \ and\ \bibinfo
  {author} {\bibfnamefont {D.}~\bibnamefont {Robinson}},\ }\bibfield
  {booktitle} {\emph {\bibinfo {booktitle} {{Proceedings, 38th International
  Conference on High Energy Physics (ICHEP 2016): Chicago, IL, USA, August
  3-10, 2016}}},\ }\href@noop {} {\bibfield  {journal} {\bibinfo  {journal}
  {PoS}\ }\textbf {\bibinfo {volume} {ICHEP2016}},\ \bibinfo {pages} {1074}
  (\bibinfo {year} {2017})}\BibitemShut {NoStop}%
\end{thebibliography}

\end{document}